\documentclass[prb,twocolumn,groupedaddress]{revtex4-1}

\usepackage{graphicx}
\usepackage{amsmath}
\usepackage{amssymb}
\usepackage{verbatim}
\usepackage[mode=math]{siunitx}
\usepackage{color}

\begin{document}


\title{Superradiance-like Electron Transport through a Quantum Dot}


\author{M. J. A. Schuetz$^{1}$, E. M. Kessler$^{1}$, J. I. Cirac$^{1}$, and G. Giedke$^{1,2}$}
\affiliation{$^{1}$Max-Planck-Institut f\"ur Quantenoptik,
Hans-Kopfermann-Str. 1,
85748 Garching, Germany}
\affiliation{$^2$ M5, Fakult\"at f\"ur Mathematik, TU M\"unchen, L.-Boltzmannstr. 1, 85748
Garching, Germany}


\date{\today}

\begin{abstract}
We theoretically show that intriguing features of coherent many-body
physics can be observed in electron transport through a quantum dot
(QD). We first derive a master equation based framework
for electron transport in the Coulomb-blockade regime which includes
hyperfine (HF) interaction with the nuclear spin ensemble
in the QD. This general tool is then used to study the leakage
current through a single QD in a transport setting. 
We find that, for an initially polarized
nuclear system, the proposed setup leads to a strong current peak, in close
analogy with superradiant emission of photons from atomic ensembles.
This effect could be observed with realistic experimental parameters
and would provide clear evidence of coherent HF dynamics of nuclear
spin ensembles in QDs.
\end{abstract}


\maketitle


\section{Introduction}

Quantum coherence is at the very heart of many intriguing phenomena
in today's nanostructures \cite{brandes04, awschalom02}. For example,
it is the essential ingredient to the understanding of the famous
Aharonov-Bohm like interference oscillations of the conductance of
metallic rings \cite{sharvin81} or the well-known conductance steps
in quasi one-dimensional wires \cite{vanWees88, wharam88}. In particular,
nonequilibrium electronic transport has emerged as a versatile
tool to gain deep insights into the coherent quantum properties of
mesoscopic solid-state devices \cite{nazarov09, datta97}. Here, with
the prospect of spintronics and applications in quantum computing,
a great deal of research has been directed towards the interplay and
feedback mechanisms between electron and nuclear spins in gate-based
semiconductor quantum dots \cite{hanson07, vanderWiel02, johnson05, jouravlev06, baugh07, petta08, inarrea07}.
Current fluctuations have been assigned to the random dynamics of
the ambient nuclear spins \cite{koppens05} and/or hysteresis effects
due to dynamic nuclear polarization \cite{koppens05, ono04, pfund07, kobayashi11}.
Spin-flip mediated transport, realized in few-electron quantum dots
in the so-called spin-blockade regime \cite{ono02}, has been shown
to exhibit long time scale oscillations and bistability as a result
of a buildup and relaxation of nuclear polarization \cite{koppens05, ono04}.
The nuclear spins are known to act collectively on the electron spin
via hyperfine interaction. In principle, this opens up an exciting
testbed for the observation of collective effects which play a remarkable
role in a wide range of many-body physics \cite{rudner07, eto04, christ07}.

In Quantum Optics, the concept of superradiance, describing the cooperative
emission of photons, is a paradigm example for a 
cooperative quantum 
effect
\cite{brandes04, dicke54, gross82}. Here, initially excited atoms
emit photons collectively as a result of the buildup and reinforcement
of strong interatomic correlations. Its most prominent feature is
an emission intensity burst in which the system radiates much faster
than an otherwise identical system of independent emitters. This phenomenon
is of fundamental importance in quantum optics and has been studied
extensively since its first prediction by Dicke in 1954 \cite{dicke54}.
Yet, in its original form the observation of optical superradiance
has turned out to be difficult due to dephasing dipole-dipole van
der Waals interactions, which suppress a coherence buildup in atomic
ensembles. 

This paper is built upon analogies between mesoscopic solid-state physics 
and Quantum Optics: the nuclear spins surrounding a QD are identified with
an atomic ensemble, individual nuclear spins corresponding to the
internal levels of a single atom and the electrons are associated with photons.
Despite some fundamental differences -- for example, electrons are fermions, whereas
photons are bosonic particles -- this analogy stimulates conjectures about the 
potential occurence of related phenomena in these two fields of physics.
Led by this line of thought, we will address the question if superradiant 
behaviour might also be observed in a solid-state environment where the role 
of photons is played by electrons. 
To this end, we analyze a gate-based semiconductor QD in the Coulomb blockade regime, obtaining 
two main results, of both experimental and theoretical relevance:
First, in analogy to superradiant emission of photons, we show how to observe superradiant emission of electrons
in a transport setting through a QD.
We demonstrate that the proposed setup, when tuned into the spin-blockade regime,
carries clear fingerprints of cooperative emission, with no van der
Waals dephasing mechanism on relevant timescales. The spin-blockade
is lifted by the HF coupling which becomes increasingly more efficient
as correlations among the nuclear spins build up. This markedly enhances
the spin-flip rate and hence the leakage current running through the QD. 
Second, we develop a general theoretical master equation framework
that describes the nuclear spin mediated transport through a single
QD. Apart from the collective effects due to the HF interaction,  
the electronic tunneling current is shown to depend on the internal state of the ambient
nuclear spins through the effective magnetic field (Overhauser field)
produced by the hyperfine interaction. 

The paper is structured as follows: 
In Sec.~\ref{sec:Executive-Summary}, we highlight our key findings and provide an intuitive picture of our basic 
ideas, allowing the reader to grasp our main results on a qualitative level. 
By defining the underlying Hamiltonian, Sec.~\ref{sec:System} 
then describes the system in a more rigorous fashion.
Next, we present the first main result of this paper in Sec.~\ref{sec:QME}:
a general master equation for electron transport through a single QD which is coherently
enhanced by the HF interaction with the ambient nuclear spins in the QD.
It features both collective effects and feedback mechanisms between the electronic and
the nuclear subsystem of the QD.   
Based on this theoretical framework, Sec.~\ref{sec:Superradiance} puts forward the
second main result, namely 
the observation of superradiant behavior in the leakage current through a QD. 
The qualitative explanations provided in Sec.~\ref{sec:Executive-Summary} should 
allow to read this part independently of the derivation given in Sec.~\ref{sec:QME}.
Sec.~\ref{sec:Analysis-and-Numerical-Results} backs up our analytical predictions with numerical simulations.
When starting from an initially polarized nuclear spin ensemble, 
the leakage current through the QD is shown to exhibit a strong peak whose relative 
height scales linearly with the number of nuclear spins, which we identify as the  
characteristic feature of superradiant behaviour. 
In Sec.~\ref{sec:Conclusion} we draw conclusions and give an outlook on future
directions of research.  

\section{Main Results \label{sec:Executive-Summary}}

\begin{figure}
\begin{center}
\includegraphics[width=\linewidth]{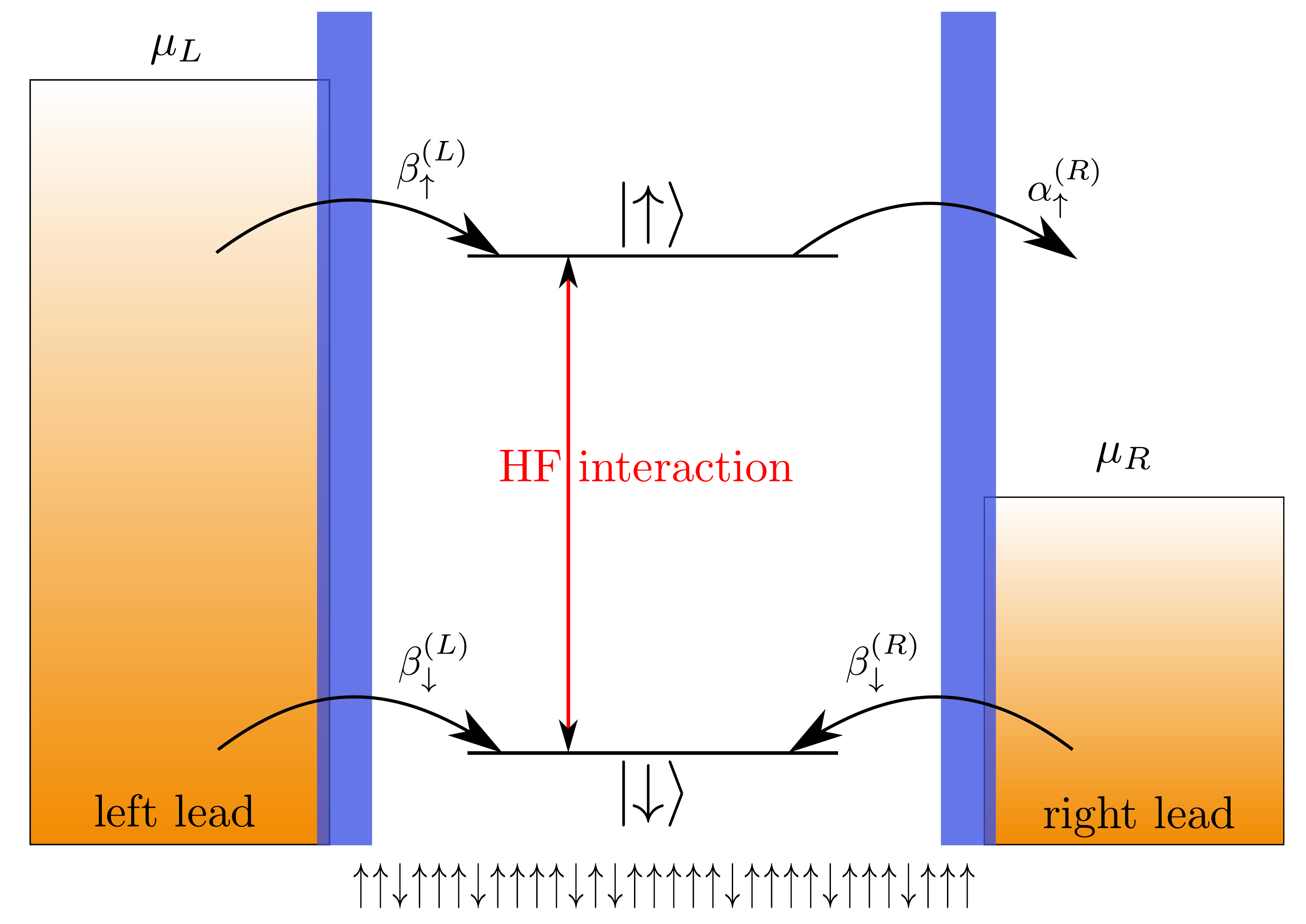}
\caption{\label{fig:system}(color online). Schematic illustration of the transport system:
An electrically defined QD is tunnel-coupled to two electron reservoirs,
the left and right lead respectively. A bias voltage $eV=\mu_{L}-\mu_{R}$
is applied between the two leads in order to induce a current through
the QD. An external magnetic field is used to tune the system into
the sequential-tunneling regime and the QD effectively acts as an spin-filter.
The resulting spin-blockade can be lifted by the HF interaction between
the QD electron and the nuclear spins in the surrounding host environment. }
\end{center}
\end{figure}

In this section we provide an intuitive
exposition of our key ideas and summarize our main findings.   
We study a single electrically-defined QD in the Coulomb-blockade regime which is 
attached to two leads, as schematically depicted in Fig.~\ref{fig:system}.
Formally, the Hamiltonian for the total system is given by 
\begin{equation}
H=H_{Z}+H_{B}+H_{T}+H_{\mathrm{HF}}.\label{eq:Hamiltonian}
\end{equation}
Here, $H_{Z}$ describes the Zeeman splitting of the electronic levels inside the QD
in presence of an external magnetic field. $H_{B}$ refers to two independent reservoirs of
non-interacting electrons, the left and right lead respectively. 
The coupling between these and the QD is described in terms of a tunneling
Hamiltonian $H_{T}$ and $H_{\mathrm{HF}}$ models the \textit{collective} interaction between an 
electron confined inside the QD and an ensemble of $N$ proximal nuclear spins surrounding the QD.  
Note that the specific form of $H$ will be given later on in Sec.~\ref{sec:System}.

Our analysis is built upon a Quantum master equation approach, a technique originally
rooted in the field of Quantum Optics. By tracing out the unobserved
degrees of freedom of the leads we derive an effective equation of motion 
for the density matrix of the QD system $\rho_{S}$ 
-- describing the electron spin inside the QD as well as the nuclear spin ensemble -- 
irreversibly coupled to source and drain electron reservoirs.
In addition to the standard
assumptions of a weak system-reservoir coupling (Born approximation), a flat reservoir spectral
density, and a short reservoir correlation time (Markov approximation), we demand the
hyperfine flip-flops to be strongly detuned with respect to the effective magnetic
field seen by the electron throughout the dynamics.
Under these conditions, the central master equation can be written as 
\begin{eqnarray}
\label{eq:QME}
\dot{\rho}_{S}\left(t\right) & = & -i\left[H_{Z}+H_{\mathrm{HF}},\rho_{S}\left(t\right)\right] \\
 &  & +\sum_{\sigma=\uparrow,\downarrow}\alpha_{\sigma}\left(t\right)\left[d_{\sigma}\rho_{S}\left(t\right)d_{\sigma}^{\dagger}-\frac{1}{2}\left\{ d_{\sigma}^{\dagger}d_{\sigma},\rho_{S}\left(t\right)\right\} \right]\nonumber \\
 &  & +\sum_{\sigma=\uparrow,\downarrow}\beta_{\sigma}\left(t\right)\left[d_{\sigma}^{\dagger}\rho_{S}\left(t\right)d_{\sigma}-\frac{1}{2}\left\{ d_{\sigma}d_{\sigma}^{\dagger},\rho_{S}\left(t\right)\right\} \right], \nonumber 
\end{eqnarray}
where the tunneling rates $\alpha_{\sigma}\left(t\right)$ and $\beta_{\sigma}\left(t\right)$
describe dissipative processes by which an electron of spin $\sigma$ tunnels from one of the 
leads into or out of the QD, respectively.   
Here, the fermionic operator $d_{\sigma}^{\dagger}$ creates an electron of spin $\sigma$
inside the QD.
While a detailed derivation of Eqn.\eqref{eq:QME} along with the precise form
of the tunneling rates is presented in Sec.~\ref{sec:QME}, here we focus on 
a qualitative discussion of the theoretical and experimental implications thereof.    

Our central master equation 
exhibits two core features:
First, 
dissipation only acts on the electronic subsystem
with rates $\alpha_{\sigma}\left(t\right)$ and $\beta_{\sigma}\left(t\right)$
that depend dynamically
on the state of the nuclear subsystem. This non-linear behavior potentially
results in hysteretic behavior and feedback mechanisms between the two subsystems 
as already suggested theoretically \cite{jouravlev06, inarrea07, rudner07, eto04} 
and observed in experiments in the context of double QDs in the Pauli-blockade regime; 
see e.g. Refs.~\cite{baugh07, petta08, kobayashi11}.
Second, the collective nature of the HF interaction $H_{\mathrm{HF}}$ allows for the
observation of coherent many-body effects.  

The effect of the hyperfine interaction between an electron inside the QD 
and the ambient nuclear spin ensemble is two-fold giving rise to the two 
main results outlined above:
First, the nuclear spins provide an effective magnetic field for the electron spin, 
the Overhauser field, whose strength is proportional to the polarization of the nuclear spin ensemble. 
Thus, a changing nuclear polarization can either dynamically tune or detune the position of the 
electron levels inside the QD. This, in turn, can have a marked effect on the transport properties of the QD
as they crucially depend on the position of these resonances with respect to 
the chemical potentials of the leads. 
In our model, this effect is directly captured by the tunneling rates dynamically depending
on the state of the nuclei.  
Second, to show that this system supports the observation of intriguing, purely collective
effects we refer to the following example:     
Consider a setting in which
the bias voltage and an external magnetic field are tuned such that only one of the
two electronic spin-components, say the level $\left|\uparrow\right\rangle $,
lies inside the transport window. In this spin-blockade regime the
electrons tunneling into the right lead are spin-polarized, i.e., the
QD acts as an spin filter \cite{recher00, hanson04}. 
If the HF coupling is sufficiently small compared to the external Zeeman splitting,
the electron is predominantly in its $\left|\downarrow\right>$ spin state allowing
to adiabatically eliminate the electronic QD coordinates. In this way we obtain
an effective equation of motion for
the nuclear density operator $\mu$ only. It reads 
\begin{eqnarray}
\dot{\mu} & = & c_{r}\left[A^{-}\mu A^{+}-\frac{1}{2}\left\{ A^{+}A^{-},\mu\right\} \right]\nonumber \\
 &  & +ic_{i}\left[A^{+}A^{-},\mu\right]+i\frac{g}{2}\left[A^{z},\mu\right],\label{eq:QME-nuclear-spins}
\end{eqnarray}
where $A^{\mu}=\sum_{i=1}^{N}g_{i}\sigma_{i}^{\mu}$
with $\mu=+,-,z$ are \textit{collective} nuclear spin operators, composed of \textit{all} $N$
individual nuclear spin operators $\sigma_{i}^{\mu}$, with $g_{i}$ being 
proportional to the probability of the electron being at the location of the
nucleus of site $i$. Again, we will highlight the core implications of Eqn.\eqref{eq:QME-nuclear-spins} 
and for a full derivation thereof, including
the definition of the effective rates $c_{r}$ and $c_{i}$, we refer to 
Sec.~\ref{sec:Superradiance}. 
Most notably, Eqn.\eqref{eq:QME-nuclear-spins} closely resembles the superradiance master equation
which has been discussed extensively in the context of atomic physics \cite{gross82}
and therefore similar effects might be expected.

\begin{figure}
\begin{center}
\includegraphics[width=\linewidth]{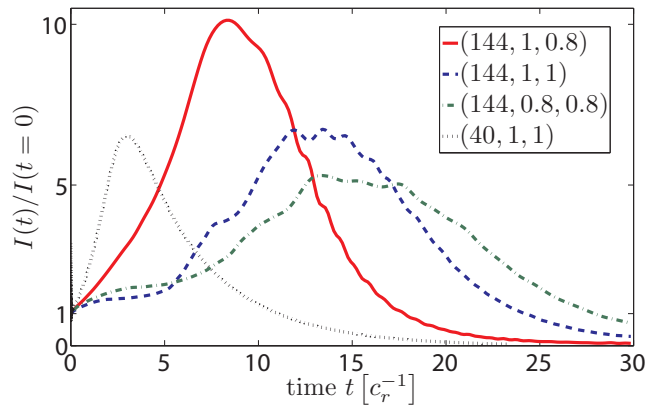}
\caption{\label{fig:SR}(color online). Normalized leakage current through a QD in the spin-blockade regime
for $N$ nuclear spins, initial nuclear polarization $p$ and external Zeeman splitting $\omega_{0}$ in units of
the total HF coupling constant $A_{\mathrm{HF}}\approx \SI{100}{\mu eV}$, summarized as 
$\left(N,p,\omega_{0}/A_{\mathrm{HF}}\right)$. For homogeneous HF coupling the dynamics can be solved exactly
(black dotted line). Compared to this idealized benchmark, the effects are reduced for realistic
inhomogeneous HF coupling, but still present: The relative peak height becomes more pronounced for smaller 
detuning $\omega_{0}$ or higher polarization $p$ (solid red line compared to the blue dashed and 
green dash-dotted line, respectively). Even under realistic conditions, the relative peak height 
is found to scale linearly with $N$, corresponding to a strong enhancement for typically $N\approx 10^5 - 10^6$.    
}
\end{center}
\end{figure}

Superradiance is known as a macroscopic collective phenomenon which generalizes spontaneous emission from a single emitter
to a many-body system of $N$ atoms \cite{brandes04}. Starting from a fully polarized initial state the system evolves within a 
totally symmetric subspace under permutation and experiences a strong correlation build-up. 
As a consequence, the emission intensity is not of the usual exponentially decaying form, but
conversely features a sudden peak occuring on a very rapid timescale $\sim1/N$ with a maximum $\sim N^2$.

In this paper, we show that the same type of cooperative emission
can occur from an ensemble of nuclear spins surrounding an electrically-defined QD, 
a phenomenon we term as \textit{electronic superradiance}:
The spin-blockade
can be lifted by the HF interaction as the nuclei pump excitations
into the electron. When starting from a highly polarized, weakly correlated 
nuclear state, due to the collective nature of the HF interaction,
this process becomes increasingly more efficient
as correlations among the nuclei build up, directly giving rise to 
an increased 
leakage current.
Therefore, the current is collectively 
enhanced by the electron's HF interaction with the ambient nuclear spin ensemble giving rise to
the pure many-body effect of electronic superradiance. 

Compared to its conventional atomic counterpart, our system incorporates two major differences:
First, our setup describes superradiant behaviour from a \textit{single} emitter, since
in the strong Coulomb-blockade regime the electrons are emitted antibunched.
As described above, the superradiant character is due to the nuclear 
spins acting collectively on the electron spin leading to an increased 
leakage current on timescales longer than single electron tunneling events.
The second crucial difference is the \textit{inhomogeneous} nature $(g_{i}\neq \mathrm{const.})$ 
of the collective operators $A^{\mu}$.
Accordingly, the collective spin is not conserved, leading to dephasing between the nuclei which
in principle could prevent the observation of superradiant behavior.
However, as exemplified in Fig.~\ref{fig:SR},
we show that under realistic conditions -- taking into account a finite initial polarization
of nuclear spins $p$ and dephasing processes due to the inhomogeneous nature of the HF coupling --
the leakage current through the QD still exhibits the characteristic peak whose
relative height will be shown to scale linealry with the number of nuclear spins. 
Even though the effect is reduced compared to the ideal atomic case,
for an experimentally realistic number of nuclei $N\approx 10^5 - 10^6$ a strong increase is still predicted. 
The experimental key signature of this effect, the relative peak height of the leakage current, can be
varied by either tuning the external Zeeman splitting or the initial polarization 
of the nuclear spins.   

In the remainder of the paper, Eqn.\eqref{eq:QME} and Eqn.\eqref{eq:QME-nuclear-spins} 
will be derived from first principles; in particular, the underlying assumptions and approximations
will be listed. Based on this general theoretical framework, more results along with detailed
discussions will be presented. For both the idealized case of homogeneous HF coupling 
-- in which an exact solution is feasible even for relatively large $N$ --
and the more realistic inhomogeneous case, further numerical simulations will prove the existence of
a strong superradiant peaking in the leakage current of single QD in the spin-blockade regime.

\section{The System \label{sec:System}}

This section gives an in-depth description of the Hamiltonian under study, 
formally introduced in Eqn.\eqref{eq:Hamiltonian}.
The system we consider consists of a single electrically-defined QD
in a transport setting as schematically depicted in Fig.~\ref{fig:system}.
Due to strong confinement only a single orbital level is relevant.
Moreover, the QD is assumed to be in the strong Coulomb-blockade regime
so that at maximum one electron resides inside the QD. Therefore,
the effective Hilbert-space of the QD electron is $\text{span}\left\{ \left|\uparrow\right\rangle ,\left|\downarrow\right\rangle ,\left|0\right\rangle \right\} $
where the lowest energy states for an additional electron in the QD
with spin $\sigma=\uparrow,\downarrow$ are split by an external magnetic
field. The Hamiltonian for the total system is given in Eqn.\eqref{eq:Hamiltonian}.
 
Here, the first term $H_{Z}=\sum_{\sigma}\epsilon_{\sigma}d_{\sigma}^{\dagger}d_{\sigma}$
describes the electronic levels of the QD. The Zeeman splitting between
the two spin components is $\omega_{0}=\epsilon_{\uparrow}-\epsilon_{\downarrow}$
(we set $\hbar=1$) and the QD electron operators are $d_{\sigma}^{\dagger}=\left|\sigma\right\rangle \left\langle 0\right|$,
describing transitions from the state $\left|0\right\rangle $ with
no electron inside the QD to a state $\left|\sigma\right\rangle $
with one electron of spin $\sigma$ inside the QD. 

Electron transport through the QD is induced by attaching the QD to
two electron leads (labeled as $L$ and $R$) which are in thermal
equilibrium at chemical potentials $\mu_{L}$ and $\mu_{R}$, respectively.
The leads themselves constitute reservoirs of non-interacting electrons
\begin{equation}
H_{B}=\sum_{\alpha,k,\sigma}\epsilon_{\alpha k}c_{\alpha k\sigma}^{\dagger}c_{\alpha k\sigma},
\end{equation}
where $c_{\alpha k\sigma}^{\dagger}\left(c_{\alpha k\sigma}\right)$
creates (annihilates) an electron in lead $\alpha=L,R$ with wavevector
$k$ and spin $\sigma$. The operators $c_{\alpha k\sigma}^{\dagger}\left(c_{\alpha k\sigma}\right)$
fulfill the usual Fermi commutation relations: $\{c_{\alpha k\sigma}^{\dagger},c_{\alpha'k'\sigma'}^{\dagger}\}=\{c_{\alpha k\sigma},c_{\alpha'k'\sigma'}\}=0$
and $\{c_{\alpha k\sigma}^{\dagger},c_{\alpha' k'\sigma'}\}=\delta_{\alpha,\alpha'}\delta_{k,k'}\delta_{\sigma,\sigma'}$.
The effect of the Coulomb interaction in the leads can be taken into
account by renormalized effective quasi-particle masses. A positive
source-drain voltage $eV=\mu_{L}-\mu_{R}$ leads to a dominant tunneling
of electrons from left to right. Microscopically, the coupling of
the QD system to the electron reservoirs is described in terms of
the tunneling Hamiltonian \begin{equation}
H_{T}=\sum_{\alpha,k,\sigma}T_{k,\sigma}^{\left(\alpha\right)}d_{\sigma}^{\dagger}c_{\alpha k\sigma}+\mathrm{h.c.},\end{equation}
with the tunnel matrix element $T_{k,\sigma}^{\left(\alpha\right)}$
specifying the transfer coupling between the lead $\alpha=L,R$ and
the system. There is no direct coupling between the leads and electron
transfer is only possible by charging and discharging the QD. 

The cooperative effects are based on the the collective hyperfine
interaction of the electronic spin of the QD with $N$ initially polarized
nuclear spins in the host environment of the QD \cite{kessler10}.
It is dominated by the isotropic contact term \cite{schliemann03}
given by \begin{equation}
H_{\mathrm{HF}}=\frac{g}{2}\left(A^{+}S^{-}+A^{-}S^{+}\right)+gA^{z}S^{z}.
\end{equation}
Here $S^{\mu}$ and $A^{\mu}=\sum_{i=1}^{N}g_{i}\sigma_{i}^{\mu}$
with $\mu=+,-,z$ denote electron and collective nuclear spin operators,
respectively. The coupling coefficients are normalized such that $\sum_{i}g_{i}^{2}=1$
and individual nuclear spin operators $\sigma_{i}^{\mu}$ are assumed
to be spin $1/2$ for simplicity; $g$ is related to the total HF
coupling strength $A_{\mathrm{HF}}$ via $g=A_{\mathrm{HF}}/\sum_{i}g_{i}$. We neglect the typically
very small nuclear Zeeman and nuclear dipole-dipole terms \cite{schliemann03}.
For simplicity, we also restrict our analysis to one nuclear species
only. These simplifications will be addressed in more detail in Sec.~\ref{sec:Analysis-and-Numerical-Results}.

The effect of the HF interaction with the nuclear spin ensemble is
two-fold: The first part of the above Hamiltonian $H_{\mathrm{ff}}=\frac{g}{2}\left(A^{+}S^{-}+A^{-}S^{+}\right)$
is a Jaynes-Cummings-type interaction which exchanges excitations
between the QD electron and the nuclei. The second term $H_{\mathrm{OH}}=gA^{z}S^{z}$
constitutes a quantum magnetic field, the Overhauser field, for the
electron spin generated by the nuclei. 
If the Overhauser field is not negligible compared to the external Zeeman splitting, it can have
a marked effect on the current by (de)tuning the hyperfine flip-flops.

\section{Generalized Quantum Master Equation \label{sec:QME}}

Electron transport through a QD can be viewed as a tool to reveal
the QD's nonequilibrium properties in terms of the current-voltage
$I/V$ characteristics. From a theoretical perspective, a great variety of
methods such as the scattering matrix formalism \cite{bruus06} and
non-equilibrium Green's functions \cite{yamamoto99, datta97} have
been used to explore the $I/V$ characteristics of quantum systems
that are attached to two metal leads. Our analysis is built upon the
master equation formalism, a tool widely used in quantum optics for
studying the irreversible dynamics of quantum systems coupled to a
macroscopic environment. 

In what follows, we will employ a projection operator based technique
to derive an effective master equation for the QD system -- comprising
the QD electron spin as well as the nuclear spins -- which experiences
dissipation via the electron's coupling to the leads. This dissipation
is shown to dynamically depend on the state of the nuclear system
potentially resulting in feedback mechanisms between the two subsystems.
We will derive conditions which allow for a Markovian treatment of
the problem and list the assumptions our master equation based framework
is based on.

\subsection{Superoperator Formalism - Nakajima-Zwanzig Equation }

The state of the global system that comprises the QD as well as the
environment is represented by the full density matrix $\rho\left(t\right)$.
However, the actual states of interest are the states of the QD which
are described by the reduced density matrix $\rho_{S}=\mathsf{Tr_{B}}\left[\rho\right]$,
where $\mathsf{Tr_{B}}\dots$ averages over the unobserved degrees
of freedom of the Fermi leads. We will derive a master equation that
governs the dynamics of the reduced density matrix $\rho_{S}$ using
the superoperator formalism. We start out from the von Neumann equation
for the full density matrix \begin{equation}
\dot{\rho}=-i\left[H\left(t\right),\rho\right],\end{equation}
where $H\left(t\right)$ can be decomposed into the following form
which turns out to be convenient later on \begin{equation}
H\left(t\right)=H_{0}\left(t\right)+H_{1}\left(t\right)+H_{T}.\end{equation}
Here, $H_{0}\left(t\right)=H_{Z}+H_{B}+g\left\langle A^{z}\right\rangle _{t}S^{z}$
comprises the Zeeman splitting caused by the external magnetic field
via $H_{Z}$ and the Hamiltonian of the non-interacting electrons
in the leads $H_{B}$; moreover, the time-dependent expectation value
of the Overhauser field has been absorbed into the definition of $H_{0}\left(t\right)$.
The HF interaction between the QD electron and the ensemble of nuclear
spins has been split up into the flip-flop term $H_{\mathrm{ff}}$
and the Overhauser field $H_{\mathrm{OH}}$, that is $H_{\mathrm{HF}}=H_{\mathrm{OH}}+H_{\mathrm{ff}}$.
The term $H_{1}\left(t\right)=H_{\Delta\mathrm{OH}}\left(t\right)+H_{\mathrm{ff}}$
comprises the Jaynes-Cummings-type dynamics $H_{\mathrm{ff}}$ and
fluctuations due to deviations of the Overhauser field from its expectation
value, i.e., $H_{\Delta\mathrm{OH}}\left(t\right)=g\delta A^{z}S^{z}$,
where $\delta A^{z}=A^{z}-\left\langle A^{z}\right\rangle _{t}$. 

The introduction of superoperators -- operators acting on the space
of linear operators on the Hilbert space -- allows for a compact notation.
The von Neumann equation is written as $\dot{\rho}=-i\mathcal{L}\left(t\right)\rho$,
where $\mathcal{L}\left(t\right)=\mathcal{L}_{0}\left(t\right)+\mathcal{L}_{1}\left(t\right)+\mathcal{L}_{T}$
is the Liouville superoperator defined via $\mathcal{L}_{\alpha}\cdot=\left[H_{\alpha},\cdot\right]$.
Next, we define the superoperator $\mathcal{P}$ as a projector onto
the relevant subspace 
\begin{equation}
\mathcal{P}\rho\left(t\right)=\mathsf{Tr_{B}}\left[\rho\left(t\right)\right]\otimes\rho_{B}^{0}=\rho_{S}\left(t\right)\otimes\rho_{B}^{0},
\end{equation}
where $\rho_{B}^{0}$ describes separate thermal equilibria of the
two leads whose chemical potentials are different due to the bias
voltage $eV=\mu_{L}-\mu_{R}$. Essentially, $\mathcal{P}$ maps a
density operator onto one of product form with the environment in
equilibrium but still retains the relevant information on the system
state. The complement of $\mathcal{P}$ is $\mathcal{Q}=1-\mathcal{P}$. 

By inserting $\mathcal{P}$ and $\mathcal{Q}$ in front of both sides
of the von Neumann equation one can derive a closed equation for the
projection $\mathcal{P}\rho\left(t\right)$, which for factorized
initial condition, where $\mathcal{Q}\rho\left(0\right)=0$, can be
rewritten in the form of the generalized Nakajima-Zwanzig master equation
\begin{eqnarray}
\frac{d}{dt}\mathcal{P}\rho & = & -i\mathcal{P}\mathcal{L}\mathcal{P}\rho\nonumber \\
 &  & -\int_{0}^{t}dt'\,\mathcal{P}\mathcal{L}\mathcal{Q}\,\hat{T}e^{-i\int_{t'}^{t}d\tau\mathcal{Q}\mathcal{L}\left(\tau\right)}\mathcal{Q}\mathcal{L}\mathcal{P}\rho\left(t'\right),
\end{eqnarray}
which is non-local in time and contains all orders of the system-leads
coupling \cite{welack08}. Here, $\hat{T}$ denotes the chronological
time-ordering operator. Since $\mathcal{P}$ and $\mathcal{Q}$ are projectors 
onto orthogonal subspaces that are only connected by $\mathcal{L}_{T}$, this
simplifies to 
\begin{equation}
\frac{d}{dt}\mathcal{P}\rho=-i\mathcal{P}\mathcal{L}\mathcal{P}\rho-\int_{0}^{t}dt'\mathcal{P}\mathcal{L}_{T}\hat{T}e^{-i\int_{t'}^{t}d\tau\mathcal{Q}\mathcal{L}\left(\tau\right)}\mathcal{L}_{T}\mathcal{P}\rho\left(t'\right).\label{eq:nakajima-zwanzig}
\end{equation}
Starting out from this exact integro-differential equation, we introduce
some approximations: In the weak coupling limit we neglect all powers
of $\mathcal{L}_{T}$ higher than two (Born approximation). Consequently,
we replace $\mathcal{L}\left(\tau\right)$ by $\mathcal{L}\left(\tau\right)-\mathcal{L}_{T}$
in the exponential of Eqn.\eqref{eq:nakajima-zwanzig}. Moreover,
we make use of the fact that the nuclear spins evolve on a time-scale
that is very slow compared to all electronic processes: In other words,
the Overhauser field is quasi-static on the timescale of single electronic
tunneling events \cite{christ07,taylor07}. That is, we replace $\left\langle A^{z}\right\rangle _{\tau}$
by $\left\langle A^{z}\right\rangle _{t}$ in the exponential of Eqn.\eqref{eq:nakajima-zwanzig}
which removes the explicit time dependence in the kernel. By taking
the trace over the reservoir and using $\mathsf{Tr_{B}}\left[\mathcal{P}\dot{\rho}\left(t\right)\right]=\dot{\rho}_{S}\left(t\right)$,
we get 
\begin{eqnarray}
\dot{\rho}_{S}\left(t\right) & = & -i\left(\mathcal{L}_{Z}+\mathcal{L}_{\mathrm{HF}}\right)\rho_{S}\left(t\right)\label{eq:nakajima-zwanzig-after-Born}\\
 &  & -\int_{0}^{t}d\tau\,\mathsf{Tr_{B}}\left(\mathcal{L}_{T}e^{-i\left[\mathcal{L}_{0}\left(t\right)+\mathcal{L}_{1}\left(t\right)\right]\tau}\mathcal{L}_{T}\mathcal{P}\rho\left(t-\tau\right)\right).\nonumber 
\end{eqnarray}
Here, we also used the relations $\mathcal{P}\mathcal{L}_{T}\mathcal{P}=0$
and $\mathcal{L}_{B}\mathcal{P}=0$ and switched the integration variable
to $\tau=t-t'$. Note that, for notational convenience, we will suppress
the explicit time-dependence of $\mathcal{L}_{0(1)}\left(t\right)$
in the following. In the next step, we iterate the Schwinger-Dyson
identity \begin{eqnarray}
e^{-i\left(\mathcal{L}_{0}+\mathcal{L}_{1}\right)\tau} & = & e^{-i\mathcal{L}_{0}\tau}\label{eq:Schwinger-Dyson}\\
 &  & -i\int_{0}^{\tau}d\tau'\, e^{-i\mathcal{L}_{0}\left(\tau-\tau'\right)}\mathcal{L}_{1}e^{-i\left(\mathcal{L}_{0}+\mathcal{L}_{1}\right)\tau'}.\nonumber \end{eqnarray}

In what follows, we will keep only the first term of this infinite
series (note that the next two leading terms are explicitly calculated
in Appendix \ref{sec:Derivation-QME}). In quantum optics, this simplification
is well known as approximation of independent rates of variation \cite{cohen-tannoudji92}.
In our setting it is valid, if $\mathcal{L}_{1}\left(t\right)$ is
small compared to $\mathcal{L}_{0}\left(t\right)$ and if the bath
correlation time $\tau_{c}$ is short compared to the HF dynamics,
$A_{\mathrm{HF}}\ll1/\tau_{c}$. Pictorially, this means that during the correlation
time $\tau_{c}$ of a tunneling event, there is not sufficient time
for the Rabi oscillation with frequency $g\lesssim A_{\mathrm{HF}}$ to occur. For
typical materials\cite{timm12}, the relaxation time $\tau_{c}$ is in the range
of $\sim \SI{d-15}{s}$ corresponding to a relaxation rate
$\Gamma_{c}=\tau_{c}^{-1}\approx \SI{d5}{\mu eV}$. Indeed, this is much
faster than all other relevant processes. In this limit, the equation
of motion for the reduced density matrix of the system simplifies
to \begin{eqnarray}
\dot{\rho}_{S}\left(t\right) & = & -i\left(\mathcal{L}_{Z}+\mathcal{L}_{\mathrm{HF}}\right)\rho_{S}\left(t\right)\label{eq:QME-non-Markovian-superoperator}\\
 &  & -\int_{0}^{t}d\tau\,\mathsf{Tr_{B}}\left(\mathcal{L}_{T}e^{-i\mathcal{L}_{0}\left(t\right)\tau}\mathcal{L}_{T}\rho_{S}\left(t-\tau\right)\otimes\rho_{B}^{0}\right).\nonumber 
\end{eqnarray}
Note, however, that this master equation is not Markovian as the rate
of change of $\rho_{S}\left(t\right)$ still depends on its past.
Conditions which allow for a Markovian treatment of the problem will
be addressed in the following.

\subsection{Markov Approximation}

Using the general relation $e^{-i\mathcal{L}_{0}\tau}\mathcal{O}=e^{-iH_{0}\tau}\mathcal{O}e^{iH_{0}\tau}$
for any operator $\mathcal{O}$, we rewrite Eqn.\eqref{eq:QME-non-Markovian-superoperator}
as 
\begin{widetext}
\begin{equation}
\dot{\rho}_{S}\left(t\right)=-i\left[H_{Z}+H_{\mathrm{HF}},\rho_{S}\left(t\right)\right]-\int_{0}^{t}d\tau\,\mathsf{Tr_{B}}\left(\left[H_{T},\left[\tilde{H}_{T}\left(\tau\right),e^{-iH_{0}\tau}\rho_{S}\left(t-\tau\right)e^{iH_{0}\tau}\otimes\rho_{B}^{0}\right]\right]\right).\label{eq:QME-non-Markovian-Hamiltonian}
\end{equation}
\end{widetext}
In accordance with the previous approximations, we will replace $e^{-iH_{0}\tau}\rho_{S}\left(t-\tau\right)e^{iH_{0}\tau}$
by $\rho_{S}\left(t\right)$ which is approximately the same since
any correction to $H_{0}$ would be of higher order in perturbation
theory \cite{timm08, harbola06}. In other words, the evolution of
$\rho_{S}\left(t-\tau\right)$ is approximated by its unperturbed
evolution which is legitimate provided that the relevant timescale
for this evolution $\tau_{c}$ is very short (Markov approximation).
This step is motivated by the typically rapid decay of the lead correlations
functions \cite{timm08}; the precise validity of this approximation
is elaborated below. In particular, this simplification disregards
dissipative effects induced by $H_{T}$ which is valid self-consistently
provided that the tunneling rates are small compared to the dynamics
generated by $H_{0}$. 

Moreover, in Eqn.\eqref{eq:QME-non-Markovian-Hamiltonian} we introduced
the tunneling Hamiltonian in the interaction picture as $\tilde{H}_{T}\left(\tau\right)=e^{-iH_{0}\tau}H_{T}e^{iH_{0}\tau}$.
For simplicity, we will only consider one lead for now and add the
terms referring to the second lead later on. Therefore, we can disregard
an additional index specifying the left or right reservoir and write
explicitly \begin{equation}
\tilde{H}_{T}\left(\tau\right)=\sum_{k,\sigma}T_{k,\sigma}e^{-i\left(\epsilon_{\sigma}\left(t\right)-\epsilon_{k}\right)\tau}d_{\sigma}^{\dagger}c_{k\sigma}+\mathrm{h.c}.\end{equation}
Here, the resonances $\epsilon_{\sigma}\left(t\right)$ are explicitly
time-dependent as they dynamically depend on the polarization of the
nuclear spins\begin{equation}
\epsilon_{\uparrow\left(\downarrow\right)}\left(t\right)=\epsilon_{\uparrow\left(\downarrow\right)}\pm\frac{g}{2}\left\langle A^{z}\right\rangle _{t}.\label{eq:resonances}\end{equation}
The quantity \begin{equation}
\omega=\epsilon_{\uparrow}\left(t\right)-\epsilon_{\downarrow}\left(t\right)=\omega_{0}+g\left\langle A^{z}\right\rangle _{t}\end{equation}
can be interpreted as an effective Zeeman splitting which incorporates
the external magnetic field as well as the mean magnetic field generated
by the nuclei. 

Since the leads are assumed to be at equilibrium, their correlation
functions are given by \begin{equation}
\mathsf{Tr_{B}}\left[c_{k\sigma}^{\dagger}\left(\tau\right)c_{k'\sigma'}\rho_{B}^{0}\right]=\delta_{\sigma,\sigma'}\delta_{k,k'}e^{-i\epsilon_{k}\tau}f_{k}\label{eq:correlatoin-function-1}\end{equation}
\begin{equation}
\mathsf{Tr_{B}}\left[c_{k\sigma}\left(\tau\right)c_{k'\sigma'}^{\dagger}\rho_{B}^{0}\right]=\delta_{\sigma,\sigma'}\delta_{k,k'}e^{i\epsilon_{k}\tau}\left(1-f_{k}\right),\label{eq:correlation-function-2}\end{equation}
where the Fermi function $f_{k}=\left(1+\exp\left[\beta\left(\epsilon_{k}-\mu\right)\right]\right)^{-1}$
with inverse temperature $\beta=1/k_{B}T$ gives the thermal occupation
number of the respective lead in equilibrium. Note that all terms
comprising two lead creation $c_{k\sigma}^{\dagger}$ or annihilation
operators $c_{k\sigma}$ vanish since $\rho_{B}^{0}$ contains states
with definite electron number only \cite{timm08}. The correlation
functions are diagonal in spin space and the tunneling Hamiltonian
preserves the spin projection; therefore only co-rotating terms prevail.
If we evaluate all dissipative terms appearing in Eqn.\eqref{eq:QME-non-Markovian-Hamiltonian},
due to the conservation of momentum and spin in Eqn.\eqref{eq:correlatoin-function-1}
and Eqn.\eqref{eq:correlation-function-2}, only a single sum over
$k,\sigma$ survives. Here, we single out one term explicitly, but
all other terms follow analogously. We obtain \begin{equation}
\dot{\rho}_{S}\left(t\right)=\ldots+\sum_{\sigma}\int_{0}^{t}d\tau\,\mathcal{C}_{\sigma}\left(\tau\right)d_{\sigma}^{\dagger}e^{-iH_{0}\tau}\rho_{S}\left(t-\tau\right)e^{iH_{0}\tau}d_{\sigma},\end{equation}
where the correlation time of the bath $\tau_{c}$ is determined by
the decay of the noise correlations \begin{eqnarray}
\mathcal{C}_{\sigma}\left(\tau\right) & = & \sum_{k}\left|T_{k,\sigma}\right|^{2}f_{k}e^{i\left(\epsilon_{\sigma}\left(t\right)-\epsilon_{k}\right)\tau}\nonumber \\
 & = & \int_{0}^{\infty}d\epsilon\, J_{\sigma}\left(\epsilon\right)e^{i\left(\epsilon_{\sigma}\left(t\right)-\epsilon\right)\tau}.\end{eqnarray}
Here, we made use of the fact that the leads are macroscopic and therefore
exhibit a continuous density of states per spin $n\left(\epsilon\right)$.
On top of that, we have introduced the spectral density of the bath
as\begin{equation}
J_{\sigma}\left(\epsilon\right)=D_{\sigma}\left(\epsilon\right)f\left(\epsilon\right),\end{equation}
where $D_{\sigma}\left(\epsilon\right)=n\left(\epsilon\right)\left|T_{\sigma}\left(\epsilon\right)\right|^{2}$
is the effective density of states. The Markovian treatment manifests
itself in a self-consistency argument: We assume that the spectral
density of the bath $J_{\sigma}\left(\epsilon\right)$ is flat around
the (time-dependent) resonance $\epsilon_{\sigma}\left(t\right)$
over a range set by the characteristic width $\Gamma_{\mathrm{d}}$. Typically,
both the tunneling matrix elements $T_{\sigma}\left(\epsilon\right)$
as well as the density of states $n\left(\epsilon\right)$ are slowly
varying functions of energy. In the so-called wide-band limit the
effective density of states $D_{\sigma}\left(\epsilon\right)$ is
assumed to be constant so that the self-consistency argument will
exclusively concern the behaviour of the Fermi function $f\left(\epsilon\right)$
which is intimately related to the temperature of the bath $T$. Under
the condition, that $J_{\sigma}\left(\epsilon\right)$ behaves flat on the scale $\Gamma_{\mathrm{d}}$, 
it can be replaced
by its value at $\epsilon_{\sigma}\left(t\right)$, and the noise
correlation simplifies to 
\begin{equation}
\mathcal{C}_{\sigma}\left(\tau\right)=J_{\sigma}\left(\epsilon_{\sigma}\left(t\right)\right)e^{i\epsilon_{\sigma}\left(t\right)\tau}\int_{0}^{\infty}d\epsilon\, e^{-i\epsilon\tau}.
\end{equation}
Using the relation 
\begin{equation}
\int_{0}^{\infty}d\epsilon\, e^{-i\epsilon\tau}=\pi\delta\left(\tau\right)-i\mathbb{P}\frac{1}{\tau},
\end{equation}
with $\mathbb{P}$ denoting Cauchy's principal value, we find that
the Markov approximation $\mathrm{Re}\left[\mathcal{C}_{\sigma}\left(\tau\right)\right]\propto\delta\left(\tau\right)$
is fulfilled provided that the self-consistency argument holds. This
corresponds to the white-noise limit where the correlation-time of
the bath is $\tau_{c}=0$. Pictorially, the reservoir has no memory
and instantaneously relaxes to equilibrium. We can then indeed replace
$e^{-iH_{0}\tau}\rho_{S}\left(t-\tau\right)e^{iH_{0}\tau}$ by $\rho_{S}\left(t\right)$
and extend the integration in Eqn.\eqref{eq:QME-non-Markovian-Hamiltonian}
to infinity, with neglibile contributions due to the rapid decay of
the memory kernel. In the following, we will derive an explicit condition
for the self-consistency argument to be satisfied. 

\begin{figure}
\begin{center}
\includegraphics[width=1.09 \linewidth]{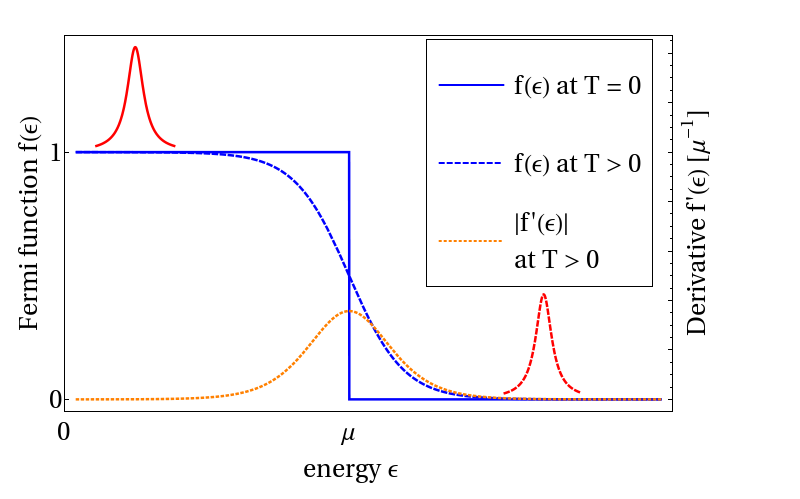}

\caption{(color online). Fermi function for finite temperature (dashed blue line) and in the limit
$T=0$ (solid blue line). The absolute value of the derivative of the Fermi function $f'(\epsilon)$ (dotted orange line for finite temperature)
is maximized at the chemical potential $\mu$ and tends to a delta function in the limit $T\rightarrow0$. 
The Markovian description is valid provided that
the Fermi function is approximately constant around the resonances $\epsilon_{\sigma}\left( t \right)$
on a scale of the width of these resonances, schematically shown in red (solid line for $\epsilon_{\sigma}\left( t \right)<\mu$
and dashed line for $\epsilon_{\sigma}\left( t \right)>\mu$). 
\label{fig:Fermi-function}}
\end{center}
\end{figure}

Let us first consider the limit $T=0$: As schematically depicted
in Fig.~\ref{fig:Fermi-function}, in this case $f\left(\epsilon\right)$
behaves perfectly flat except for $\epsilon=\mu$ where the self-consisteny
argument is violated. Therefore, the Markovian approximation is valid
at $T=0$ given that the condition $\left|\epsilon_{\sigma}\left(t\right)-\mu\right|\gg\Gamma_{\mathrm{d}}$
is fulfilled. In this limit, all tunneling rates are constant over
time and effectively decoupled from the nuclear dynamics. Note that
for the observation of electronic superradiance it will be sufficient
to restrict oneself to this case. 

For a more general analysis, we now turn to the case of finite temperature
$T>0$. We require the absolute value of the relative change of
the Fermi function around the resonance $\epsilon_{\sigma}\left(t\right)$
over a range of the characteristic width $\Gamma_{\mathrm{d}}$ to
be much less than unity, that is \begin{equation}
\left|\left.\frac{\partial f\left(\epsilon\right)}{\partial\epsilon}\right|_{\epsilon_{\sigma}\left(t\right)}\right|\Gamma_{\mathrm{d}}\ll1.\end{equation}
An upper bound for the first factor can easily be obtained as this
quantity is maximized at the chemical potential $\mu$, for all temperatures.
Evaluating the derivative at $\epsilon_{\sigma}\left(t\right)=\mu$
results in the compact condition\begin{equation}
\Gamma_{\mathrm{d}}\ll4k_{B}T.\label{eq:self-consistency-condition}\end{equation}
Thus, finite temperature $T>0$ washes out the rapid character of
$f\left(\epsilon\right)$ at the chemical potential $\mu$ and, provided
that Eqn.\eqref{eq:self-consistency-condition} is fulfilled, allows
for an Markovian treatment. 

Two distinct mechanisms contribute to
the width $\Gamma_{\mathrm{d}}$: dissipation due to coupling to the
leads and the effect of $H_{1}\left(t\right)$, because both of them
have been neglected self-consistently in the memory kernel when going
from Eqn.\eqref{eq:nakajima-zwanzig} to Eqn.\eqref{eq:QME-non-Markovian-superoperator}.
Typically, the tunneling rates are of the order of $\sim 5-20\SI{}{\mu eV}$,
depending on the transparency of the tunnel-barrier. Regarding the
contribution due to $H_{1}\left(t\right)$, we first consider two
limits of particular importance: For a completely mixed state the
fluctuation of the nuclear field around its zero expectation value
is of the order of $\sim A_{\mathrm{HF}}/\sqrt{N}\approx \SI{0.1}{\mu eV}$. In contrast,
for a fully polarized state these fluctuations can be neglected whereas
the effective strength of the flip-flop dynamics is $\sim A_{\mathrm{HF}}/\sqrt{N}$
as well. Therefore, in both limits considered here, the dominant contribution
to $\Gamma_{\mathrm{d}}$ is due to the coupling to the leads and
the self-consistency condition could still be met with cryostatic
temperatures $k_{B}T\gtrsim \SI{10}{\mu eV}$, well below the orbital level
spacing. However, we note that in the course of a superradiant evolution,
where strong correlations among the nuclei build up, the dominant
contribution to $\Gamma_{\mathrm{d}}$ may come from the flip-flop
dynamics, which are $A_{\mathrm{HF}}/4\approx \SI{25}{\mu eV}$ at maximum for homogeneous coupling. 
For realisitic conditions, though, this effect is significantly reduced,
as will be demonstrated in our simulations in Sec.~\ref{sec:Analysis-and-Numerical-Results}.

\subsection{General Master Equation for Nuclear Spin Assisted Transport \label{sec:Central-QME}}

Assuming that the self-consistency argument for a Markovian treatment
is satisfied, we now apply the following modifications to Eqn.\eqref{eq:QME-non-Markovian-Hamiltonian}:
First, we neglect level shifts due to the coupling to the continuum
states which can be incorporated by replacing the bare frequencies
$\epsilon_{\sigma}\left(t\right)$ with renormalized frequencies.
Second, one adds the second electron reservoir that has been omitted
in the derivation above. Lastly, one performs a suitable transformation
into a frame rotating at the frequency $\bar{\epsilon}=\left(\epsilon_{\uparrow}+\epsilon_{\downarrow}\right)/2$
leaving all terms invariant but changing $H_{Z}$ from $H_{Z}=\epsilon_{\uparrow}d_{\uparrow}^{\dagger}d_{\uparrow}+\epsilon_{\downarrow}d_{\downarrow}^{\dagger}d_{\downarrow}$
to $H_{Z}=\omega_{0}S^{z}$. 
After these manipulations one arrives at the central master equation as stated in Eqn.\eqref{eq:QME} 
where the tunneling rates with $\alpha_{\sigma}\left(t\right)=\sum_{x=L,R}\alpha_{\sigma}^{\left(x\right)}\left(t\right)$,
$\beta_{\sigma}\left(t\right)=\sum_{x=L,R}\beta_{\sigma}^{\left(x\right)}\left(t\right)$
and 
\begin{eqnarray*}
\frac{\alpha_{\sigma}^{\left(x\right)}\left(t\right)}{2\pi} & = & n_{x}\left(\epsilon_{\sigma}\left(t\right)\right)\left|T_{\sigma}^{\left(x\right)}\left(\epsilon_{\sigma}\left(t\right)\right)\right|^{2}\left[1-f_{x}\left(\epsilon_{\sigma}\left(t\right)\right)\right]\\
\frac{\beta_{\sigma}^{\left(x\right)}\left(t\right)}{2\pi} & = & n_{x}\left(\epsilon_{\sigma}\left(t\right)\right)\left|T_{\sigma}^{\left(x\right)}\left(\epsilon_{\sigma}\left(t\right)\right)\right|^{2}f_{x}\left(\epsilon_{\sigma}\left(t\right)\right)
\end{eqnarray*}
govern the dissipative processes in which the QD system exchanges
single electrons with the leads. The tunneling rates, as presented here, are widely
used in nanostructure quantum transport problems \cite{engel02,timm08,zhao11}.
However, in our setting they are evaluated at the resonances $\epsilon_{\sigma}\left(t\right)$
which dynamically depend on the polarization of the nuclear spins;
see Eqn.\eqref{eq:resonances}. Note that Eqn.\eqref{eq:QME}
incorporates finite temperature effects via the Fermi functions of
the leads. This potentially gives rise to feedback mechansims between
the electronic and the nuclear dynamics, since the purely electronic
diffusion markedly depends on the nuclear dynamics.

Since Eqn.\eqref{eq:QME} marks our first main result, at this
point we quickly reiterate the assumptions our master equation treatment
is based on: 
\begin{itemize}
\item The system-lead coupling is assumed to be weak and therefore treated
perturbatively up to second order (Born-approximation).
\item In particular, the tunneling rates are small compared to the effective Zeeman splitting
$\omega$.  
\item Level shifts arising from the coupling to the continuum states in
the leads are merely incorporated into a redefinition of the QD energy
levels $\epsilon_{\sigma}\left(t\right)$.
\item There is a separation of timescales between electron-spin dynamics
and nuclear-spin dynamics. In particular, the Overhauser field $g\left\langle A^{z}\right\rangle _{t}$
evolves on a timescale that is slow compared to single electron tunneling
events. 
\item We have applied the approximation of independent rates of variation:
If the HF dynamics generated by $H_{1}\left(t\right)=H_{\mathrm{ff}}+H_{\Delta\mathrm{OH}}\left(t\right)$
is (i) sufficiently weak compared to $H_{0}$ and (ii) slow compared to
the correlation time of the bath $\tau_{c}$, their effect in
the memory kernel of the master equation can be neglected. The latter
holds for $A_{\mathrm{HF}}\tau_{c}\ll1$. Note that the flip-flop dynamics can become
very fast as correlations among the nuclei build up culminating in
a maximum coupling strength of $A_{\mathrm{HF}}/4$ for homogeneous coupling. 
This
potentially drives the system into the strong coupling regime where
condition (i), that is $\omega\gg||H_{1}\left(t\right)||$, might be violated. However, under realistic conditions
of inhomogeneous coupling this effect is significantly reduced. 
\item The effective density of states $D_{\sigma}\left(\epsilon\right)=n\left(\epsilon\right)\left|T_{\sigma}\left(\epsilon\right)\right|^{2}$
is weakly energy-dependent (wide-band limit). In particular, it is
flat on a scale of the characteristic widths of the resonances. 
\item The Markovian description is valid provided that either the resonances
are far away from the chemical potentials of the leads on a scale
set by the characteristic widths of the resonances or the temperature
is sufficiently high to smooth out the rapid character of the Fermi functions
of the leads. This condition is quantified in Eqn.\eqref{eq:self-consistency-condition}.
\end{itemize}
In summary, we have derived a Quantum master equation describing electronic
transport through a single QD which is collectively enhanced due to
the interaction with a large ancilla system, namely the
nuclear spin ensemble in the host environment. Eqn.\eqref{eq:QME}
incorporates two major intriguing features both of theoretical and
experimental relevance: Due to a separation of timescales, only the electronic
subsystem experiences dissipation with rates that depend dynamically
on the state of the ancilla system. This non-linearity gives rise to
feedback mechanisms between the two subsystems as well as hysteretic behavior. 
Moreover, the collective nature of the HF interaction offers the possibility to 
observe intriguing coherent many-body effects. Here, one particular outcome
is the occurence of electronic superradiance, as will
be shown in the remainder of this paper. 

Note that in the absence of HF interaction between the QD electron
and the proximal nuclear spins, i.e., in the limit $g\rightarrow0$,
our results agree with previous theoretical studies \cite{harbola06}.

\section{Electronic Superradiance \label{sec:Superradiance}}

Proceeding from our general theory derived above, this section is
devoted to the prediction and analysis of superradiant behavior of
electrons tunneling through a single QD in the Coulomb-blockade
regime; see Fig.~\ref{fig:system} for the scheme of the setup. 

We note that, in principle, an enhancement seen in the leakage current
could simply arise from the Overhauser field dynamically tuning the hyperfine flip-flops.  
However, we can still ensure that the measured change in the leakage
current through the QD is due to cooperative emission only by dynamically
compensating the Overhauser field. This can be achieved by applying
a time dependent magnetic or spin-dependent AC Stark field such that
$H_{\mathrm{comp}}\left(t\right)=-g\left\langle A^{z}\right\rangle _{t}S^{z}$
which will be done in most of our simulations below to clearly prove the existence
of superradiant behaviour in this setting.
Consequently, in our previous analysis $H_{0}\left(t\right)$ is replaced
by $H_{0}=H_{0}\left(t\right)-g\left\langle A^{z}\right\rangle _{t}S^{z}=H_{Z}+H_{B}$
so that the polarization dependence of the tunneling rates is removed
and we can drop the explicit time-dependence of the resonances $\epsilon_{\sigma}\left(t\right)\rightarrow\epsilon_{\sigma}$.
Under this condition, the master equation for the reduced system density
operator can be written as
\begin{eqnarray}
\dot{\rho}_{S}\left(t\right) & = & -i\left[\omega_{0}S^{z}+H_{\mathrm{HF}}+H_{\mathrm{comp}}\left(t\right),\rho_{S}\left(t\right)\right] \label{eq:full-QME-OHC} \\
 &  & +\sum_{\sigma=\uparrow,\downarrow}\alpha_{\sigma}\left[d_{\sigma}\rho_{S}\left(t\right)d_{\sigma}^{\dagger}-\frac{1}{2}\left\{ d_{\sigma}^{\dagger}d_{\sigma},\rho_{S}\left(t\right)\right\} \right]\nonumber \\
 &  & +\sum_{\sigma=\uparrow,\downarrow}\beta_{\sigma}\left[d_{\sigma}^{\dagger}\rho_{S}\left(t\right)d_{\sigma}-\frac{1}{2}\left\{ d_{\sigma}d_{\sigma}^{\dagger},\rho_{S}\left(t\right)\right\} \right].\nonumber
\end{eqnarray}

In accordance with our previous considerations, in this specific setting
the Markovian treatment is valid provided that the spectral density
of the reservoirs varies smoothly around the (time-independent) resonances
$\epsilon_{\sigma}$ on a scale set by the natural widths of the level
and the fluctuations of the dynamically compensated Overhauser field.
More specifically, throughout the whole evolution the levels are assumed
to be far away from the chemical potentials of the reservoirs \cite{gurvitz96, gurvitz98}; for an illustration see Fig.~\ref{fig:Fermi-function}.
In this wide band limit, the tunneling rates $\alpha_{\sigma},\,\beta_{\sigma}$
are independent of the state of the nuclear spins. The master equation
is of Lindblad form which guarantees to preserve the complete positivity and
the hermiticity of the density matrix. Eqn.\eqref{eq:full-QME-OHC}
agrees with previous theoretical results \cite{harbola06} except
for the appearance of the collective HF interaction between the QD
electron and the ancilla system in the Hamiltonian dynamics of Eqn.\eqref{eq:full-QME-OHC}. 

To some extent, Eqn.\eqref{eq:full-QME-OHC} bears some similarity with
the quantum theory of the laser. While in the latter
the atoms interact with bosonic reservoirs, in our transport
setting the QD is pumped by the nuclear spin ensemble and emits fermionic
particles \cite{yamamoto99, zhao11}.

\begin{figure}
\begin{center}
\includegraphics[width=\linewidth]{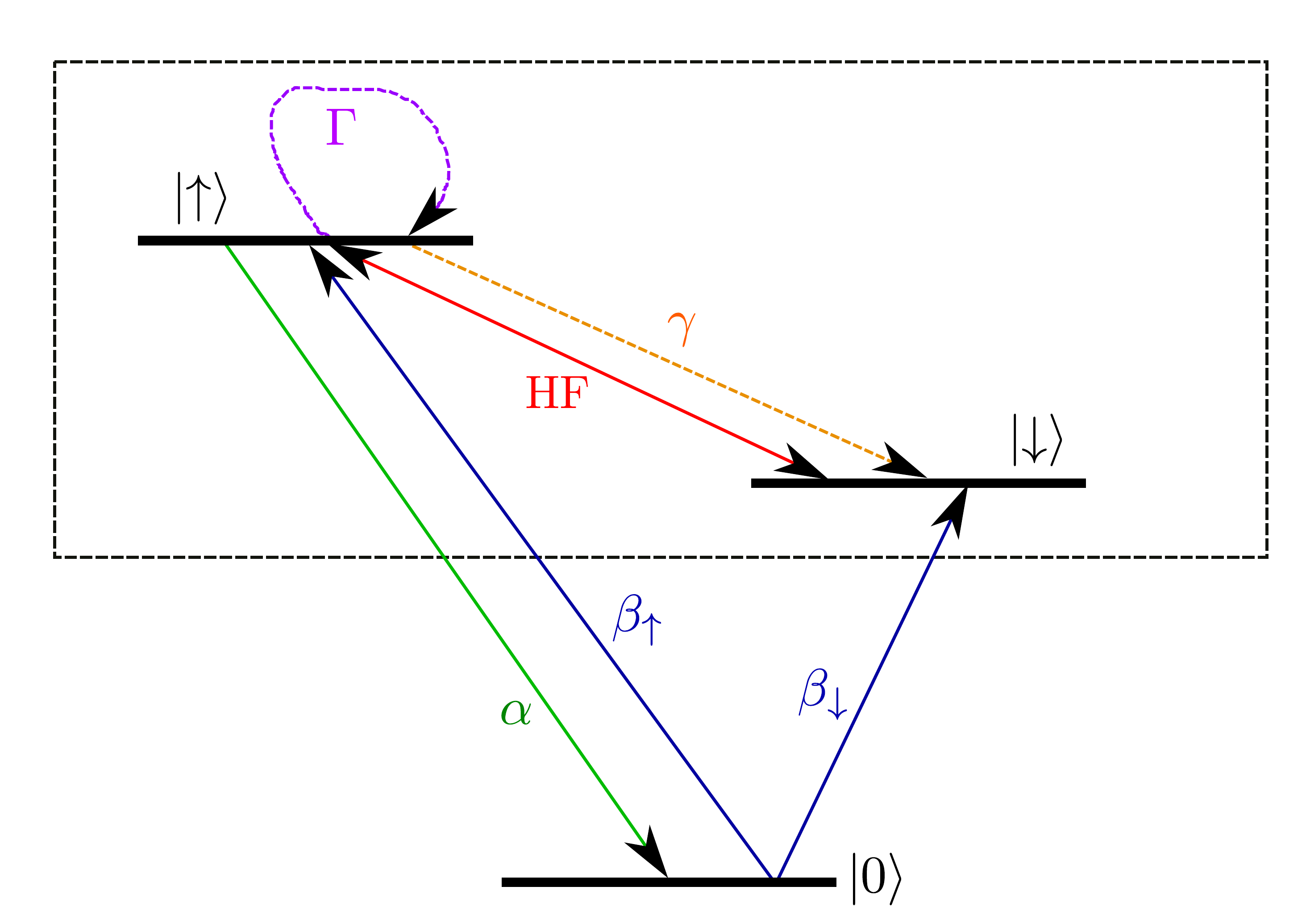}

\caption{(color online). The electronic QD system in the local moment regime after the adiabatic
elimination of the $\left|0\right\rangle $ level including the relevant
dissipative processes. Within the effective system (box) we
encounter an effective decay term and an effective pure dephasing
term, given by the rates $\gamma$ and $\Gamma$ respectively. This
simplification is possible for fast recharging of the QD, i.e., $\beta\gg\alpha$.
\label{fig:The-system-after-elimination}}
\end{center}
\end{figure}

If the HF dynamics are the slowest timescale in the problem, Eqn.\eqref{eq:full-QME-OHC}
can be recast into a form which makes its superradiant character more
apparent. In this case, the system is subject to the slaving principle
\cite{yamamoto99}: The dynamics of the whole system follow that of
the subsystem with the slowest time constant allowing to adiabatically
eliminate the electronic QD coordinates and to obtain an effective
equation of motion for the nuclear spins. In this limit, the Overhauser
field is much smaller than the Zeeman splitting so that a dynamic
compensation of the OH can be disregarded for the moment. For simplicity
we consider a transport setting in which only four tunneling rates
are different from zero, see Fig.~\ref{fig:system}. The QD can be
recharged from the left and the right lead, but only electrons with
spin projection $\sigma=\uparrow$ can tunnel out of the QD into the
right lead. We define the total recharging rate $\beta=\beta_{\downarrow}+\beta_{\uparrow}=\beta_{\downarrow}^{\left(L\right)}+\beta_{\downarrow}^{\left(R\right)}+\beta_{\uparrow}^{\left(L\right)}$
and for notational convenience unambiguously set $\alpha=\alpha_{\uparrow}^{\left(R\right)}$.
First, we project Eqn.\eqref{eq:full-QME-OHC} onto the populations of
the electronic levels and the coherences in spin space according to $\rho_{mn}=\left<m\right|\rho_{S}\left|n\right>$, 
where $m,n=0,\uparrow,\downarrow$. 
This yields
\begin{widetext}
\begin{eqnarray}
\dot{\rho}_{00} & = & \alpha\rho_{\uparrow\uparrow}-\beta\rho_{00}\\
\dot{\rho}_{\uparrow\uparrow} & = & -i\frac{g}{2}\left[A^{z},\rho_{\uparrow\uparrow}\right]-i\frac{g}{2}\left(A^{-}\rho_{\downarrow\uparrow}-\rho_{\uparrow\downarrow}A^{+}\right)-\alpha\rho_{\uparrow\uparrow}+\beta_{\uparrow}\rho_{00}\\
\dot{\rho}_{\downarrow\downarrow} & = & +i\frac{g}{2}\left[A^{z},\rho_{\downarrow\downarrow}\right]-i\frac{g}{2}\left(A^{+}\rho_{\uparrow\downarrow}-\rho_{\downarrow\uparrow}A^{-}\right)+\beta_{\downarrow}\rho_{00}\\
\dot{\rho}_{\uparrow\downarrow} & = & -i\omega_{0}\rho_{\uparrow\downarrow}-i\frac{g}{2}\left(A^{z}\rho_{\uparrow\downarrow}+\rho_{\uparrow\downarrow}A^{z}\right)-i\frac{g}{2}\left(A^{-}\rho_{\downarrow\downarrow}-\rho_{\uparrow\uparrow}A^{-}\right)-\frac{\alpha}{2}\rho_{\uparrow\downarrow}.
\end{eqnarray}
\end{widetext}
We can retrieve an effective master equation for the 
regime in which on relevant timescales the QD is always populated
by an electron. This holds for a sufficiently strong recharging rate,
that is in the limit $\beta\gg\alpha$, which can be implemented experimentally
by making the left tunnel barrier more transparent than the right
one. Then, the state $\left|0\right\rangle $ is populated negligibly
throughout the dynamics and can be eliminated adiabatically according
to $\rho_{00}\approx\frac{\alpha}{\beta}\rho_{\uparrow\uparrow}$.
In analogy to the Anderson impurity model, in the following this limit will 
be referred to as \textit{local moment regime}. 
The resulting effective master equation
reads 
\begin{eqnarray}
\dot{\rho}_{S} & = & -i\left[\omega_{0}S^{z}+H_{\mathrm{HF}},\rho_{S}\right]\label{eq:QME-local-moment}\\
 &  & +\gamma\left[S^{-}\rho_{S}S^{+}-\frac{1}{2}\left\{ S^{+}S^{-},\rho_{S}\right\} \right] \nonumber \\
 &  & +\Gamma\left[S^{z}\rho_{S}S^{z}-\frac{1}{4}\rho_{S}\right],\nonumber 
\end{eqnarray}
where $\gamma=\frac{\beta_{\downarrow}}{\beta}\alpha$ is an effective
decay rate and $\Gamma=\frac{\beta_{\uparrow}}{\beta}\alpha$ an effective
dephasing rate. This situation is schematized in Fig.~\ref{fig:The-system-after-elimination}.
The effective decay (dephasing) describes processes in which the QD
is recharged with a spin down (up) electron after a spin up electron
has tunneled out of the QD. 

In the next step we aim for an effective description that contains
only the nuclear spins: Starting from a fully polarized state, SR
is due to the increase in the operative HF matrix element $\left\langle A^{+}A^{-}\right\rangle $.
The scale of the coupling is set by the total HF coupling constant
$A_{\mathrm{HF}}=g\sum_{i}g_{i}$. For a sufficiently small \textit{relative coupling strength} \cite{kessler10}
\begin{equation}
\epsilon=A_{\mathrm{HF}}/\left(2\Delta\right),
\end{equation}
where 
\begin{equation}
\Delta=\left|\alpha/2+i\omega_{0}\right|,
\end{equation}
the electron is predominantly in its $\left|\downarrow\right\rangle $ spin state
and we can project Eqn.\eqref{eq:QME-local-moment} to the respective
subspace. As shown in detail in Appendix \ref{sec:Adiabatic-Elimination},
in this limit the reduced master equation for the nuclear density
operator $\mu=\mathsf{Tr_{el}}\left[\rho_{S}\right]$ is given by Eqn.\eqref{eq:QME-nuclear-spins},
where the effective coefficients read
\begin{eqnarray}
c_{r} & = & \frac{g^{2}\alpha}{4 \Delta^{2}},\\
c_{i} & = & \frac{g^{2}\omega_{0}}{4 \Delta^{2}}.
\end{eqnarray}
This master equation is our second main result. In an optical setting,
it has previously been predicted theoretically to exhibit strong SR signatures \cite{kessler10}.
Conceptually, its superradiant character can be understood immediately
in the ideal case of homogeneous coupling in which the collective
state of all nuclear spins can be described in terms of Dicke states
$\left|J,m\right\rangle $: The enhancement of the HF interaction
is directly associated with the transition through nuclear Dicke states
$\left|J,m\right\rangle $, $m\ll J$. In this idealized setting,
the angular momentum operator $\mathbf{I}=\sqrt{N}\mathbf{A}$ of the nuclear spin ensemble
obeys the $\text{SU(2)}$ Lie algebra, from which one can deduce the
ladder operator relation $I^{-}\left|J,m\right\rangle =\sqrt{J(J+1)-m\left(m-1\right)}\left|J,m-1\right\rangle $.
This means that, starting from an initially fully polarized
state $\left|J=N/2,m=N/2\right\rangle$, the system cascades down the Dicke-ladder with an effective
rate 
\begin{equation}\label{eq:effective-rate}
\Gamma_{m\rightarrow m-1}=\frac{c_{r}}{N}\left(N/2+m\right)\left(N/2-m+1\right),
\end{equation}
since, according to the first term in Eqn.\eqref{eq:QME-nuclear-spins},
the populations of the Dicke states evolve as 
\begin{eqnarray}
\dot{\mu}_{m,m} & = & -\frac{c_{r}}{N}\left(N/2+m\right)\left(N/2-m+1\right)\mu_{m,m} \\
 &  & +\frac{c_{r}}{N}\left(N/2+m+1\right)\left(N/2-m\right)\mu_{m+1,m+1}. \nonumber
\end{eqnarray}
While the effective rate is $\Gamma_{N/2\rightarrow N/2-1}=c_{r}$ at
the very top of the the ladder it increases up to $\Gamma_{\left|m\right|\ll N/2}\approx \frac{c_{r}}{4} N$
at the center of the Dicke ladder. This implies the characteristic
intensity peaking as compared to the limit of independent classical
emitters the emission rate of which would be $\Gamma_{\mathrm{cl}} = \frac{c_{r}}{N}N_{\uparrow}=\frac{c_{r}}{N}\left(N/2+m\right)$. 

However, there is also a major difference compared to the superradiant 
emission of photons from atomic ensembles: In contrast to its atomic
cousin, the prefactor $c_{r}/N \propto 1/N^{2}$ is $N$-dependent, 
resulting in an overall time of the SR evolution $\left<t_{D}\right>$ which 
increases with $N$. By linearizing Eqn.\eqref{eq:effective-rate} for the
beginning of the superradiant evolution \cite{gross82} as 
$\Gamma_{m\rightarrow m-1} \approx c_{r} (s+1)$, where $s=N/2-m$ gives the number
of nuclear flips, one finds that the first flip takes place in an average time
$c_{r}^{-1}$, the second one in a time $(2c_{r})^{-1}$ and so on. The summation
of all these elementary time intervals gives an upper bound estimate for
the process duration till the SR peaking as
\begin{eqnarray}\label{eq:process-duration}
 \left<t_{D}\right> \lesssim \frac{2}{c_{r}} \left[1+\frac{1}{2}+\dots+\frac{1}{N/2}\right] \approx \frac{2\ln(N/2)}{c_{r}},
\end{eqnarray}
which, indeed, increases with the number of emitters as $\sim N \ln(N)$, whereas 
one obtains $\left<t_{D}\right> \sim \ln(N) / N$ 
for ordinary superradiance \cite{gross82}. Accordingly, in our 
solid-state system the characteristic 
SR peak appears at later times for higher $N$.
The underlying reason for this difference is that in the atomic setting 
each new emitter adds to the overall coupling strength, whereas
in the central spin setting a fixed overall coupling strength $A_{\mathrm{HF}}$
is distributed over an increasing number of particles. 
Note that in an actual experimental setting $N$ is not a parameter, of course.
For our theoretical discussion, though, it is convenient to fix the total HF coupling 
strength $A_{\mathrm{HF}}$ and to extrapolate from our findings to an experimentally 
relevant number of nulear spins $N$.

For large relative coupling strength $\epsilon\gg1$ the QD electron
saturates and superradiant emission is capped by the decay rate $\alpha/2$,
prohibiting the observation of a strong intensity peak. In order to circumvent
this bottleneck regime, one has to choose a detuning $\omega_{0}$
such that $0<\epsilon\leq1$. However, to realize the spin-blockade
regime, where the upper spin manifold is energetically well separated
from the lower spin manifold, the Zeeman splitting has to be of the
order of $\omega_{0}\sim A_{\mathrm{HF}}$ which guarantees $\epsilon<1$. 
In this parameter range, the early stage of the evolution
-- in which the correlation buildup necessary for SR takes place \cite{gross82}--
is well described by Eqn.\eqref{eq:QME-nuclear-spins}. 

In reality, the inhomogeneous nature $\left(g_{i}\neq\text{const}\right)$
of the collective operators $A^{\mu}$ leads to dephasing between
the nuclei, possibly preventing the phased emission necessary for
the observation of SR \cite{gross82, kessler10, agrawal71, leonardi82}.
The inhomogeneous part of last term in Eqn.\eqref{eq:QME-nuclear-spins}
-- the electron's Knight field -- causes dephasing \cite{temnov05} $\propto g\sqrt{\text{Var}\left(g_{i}\right)}/2$, 
possibly leading to symmetry reducing transitions $J\rightarrow J-1$.
Still, it has been shown that SR is also present in realistic inhomogeneous
systems \cite{kessler10}, since the system evolves in a many-body
protected manifold (MPM): The second term in Eqn.\eqref{eq:QME-nuclear-spins}
energetically separates different total nuclear spin-$J$ manifolds
protecting the correlation build-up for large enough $\epsilon$. 

The superradiant character of Eqn.\eqref{eq:QME-nuclear-spins} suggests
the observation of its prominent intensity peak in the leakage current
through the QD in the spin-blockade regime. We have employed the method
of Full-Counting-Statistics (FCS) \cite{bagrets03, esposito09} in
order to obtain an expression for the current and find (setting the
electron's charge $e=1$) 
\begin{equation}
I\left(t\right)=\alpha\rho_{\uparrow\uparrow}-\beta_{\downarrow}^{\left(R\right)}\rho_{00}.
\end{equation}
This result is in agreement with previous theoretical findings: The
current through the device is completely determined by the occupation
of the levels adjacent to one of the leads \cite{engel02,gurvitz96,bruus06}.
The first term describes the accumulation of electrons with spin $\sigma=\uparrow$
in the right lead, whereas the second term describes electrons with
$\sigma=\downarrow$ tunneling from the right lead into the QD. As
done before \cite{kessler10}, we take the ratio of the maximum current
to the initial current (the maximum for independent emitters) $I_{\mathrm{coop}}/I_{\mathrm{ind}}$
as our figure of merit: a relative intensity peak height $I_{\mathrm{coop}}/I_{\mathrm{ind}}>1$
indicates cooperative effects. One of the characteristic features
of SR is that this quantity scales linearly with the number of spins
$N$. 

In the local-moment regime, described by Eqn.\eqref{eq:QME-local-moment},
the expression for the current simplifies to $I\left(t\right)=(1-\beta_{\downarrow}^{(R)}/\beta)\alpha\left\langle S^{+}S^{-}\right\rangle _{t}\propto\left\langle S^{+}S^{-}\right\rangle _{t}$
showing that it is directly proportional to the electron inversion.
This, in turn, increases as the nuclear system pumps excitations into
the electronic system. A compact expression for the relation between
the current and the dynamics of the nuclear system can be obtained
immediately in the case of homogeneous coupling 
\begin{equation}
\frac{d}{dt}\left\langle S^{+}S^{-}\right\rangle _{t}=-\frac{d}{dt}\left\langle I^{z}\right\rangle _{t}-\gamma\left\langle S^{+}S^{-}\right\rangle _{t}.
\end{equation}
Since the nuclear dynamics are in general much slower than the electron's
dynamics, the approximate solution of this equation is $\left\langle S^{+}S^{-}\right\rangle _{t}\approx-\frac{d}{dt}\left\langle I^{z}\right\rangle _{t}/\gamma$.
As a consequence, the current $I\left(t\right)$ is proportional to
the time-derivative of the nuclear polarization
\begin{equation}
I\left(t\right)\propto-\frac{d}{dt}\left\langle I^{z}\right\rangle _{t}.
\end{equation}

Still, no matter how strong the cooperative effects are, on a timescale
of single electron tunneling events, the electrons will always be
emitted antibunched, since in the strong Coulomb-blockade regime the
QD acts as a single-electron emitter \cite{emary12}. Typically, the
rate for single-electron emission events is even below the single
tunneling rate $\alpha$ due to the spin-blockade. On electronic timescales
$\sim1/\alpha$, the SR mechanism manifests in lifting this blockade; as argued above,
the efficiency of this process is significantly enhanced by collective effects. 

Before we proceed with an in-depth analysis of the current $I(t)$,
we note that an intriguing extension of the present work would be the study
of fluctuations thereof. Insights into the statistics of the current
could be obtained by analyzing two-time correlation functions such as
$\left<n_{\uparrow}(t+\tau)n_{\uparrow}(t)\right>$, where 
$n_{\uparrow}=d^{\dagger}_{\uparrow}d_{\uparrow}$. This can conveniently be done
via the Quantum Regression Theorem \cite{carmichael99} which yields
the formal result
$\left<n_{\uparrow}(t+\tau)n_{\uparrow}(t)\right>=\mathsf{Tr_{S}}\left[n_{\uparrow}e^{\mathcal{W}\tau}\left(n_{\uparrow}\rho_{S}(t)\right)\right]$.
Here, $\mathcal{W}$ denotes the Liouvillian governing the system's dynamics 
according to $\dot{\rho}_{S}=\mathcal{W}\rho_{S}$ and $\mathsf{Tr_{S}}\left[\dots\right]$ 
refers to the trace over the system's degree of freedoms.
This procedure can be generalized to higher order correlation functions and
full evaluation of the current statistics might reveal potential connections between 
current fluctuations and cooperative nuclear dynamics.

\section{Analysis and Numerical Results \label{sec:Analysis-and-Numerical-Results}}

\subsection{Experimental Realization}

The proposed setup described here may be realized with state-of-the-art
experimental techniques. First, the Markovian regime, valid for sufficiently
large bias $eV$, is realized if the Fermi functions of the leads
are smooth on a scale set by the natural widths of the levels and
residual fluctuations due to the dynamically compensated Overhauser
field. Since for typical materials the hyperfine coupling constant is $A_{\mathrm{HF}}=1-100\SI{}{\mu eV}$
\cite{hanson07} and tunneling rates are typically of the order of $\sim \SI{10}{\mu eV}$
\cite{vanderWiel02}, this does not put a severe restriction on the
bias voltage which are routinely in the range of hundreds of $\SI{}{\mu V}$
or $\SI{}{mV}$ \cite{kobayashi11, ono02}. Second, in order to tune the
system into the spin-blockade regime, a sufficiently large external
magnetic field has to be applied. More precisely, the corresponding
Zeeman splitting $\omega_{0}$ energetically separates the upper and
lower manifolds in such a way that the Fermi function of the right
lead drops from one at the lower manifold to zero at the upper manifold.
Temperature smeares out the Fermi function around the chemical potential 
by approximately $\sim k_{B}T$. Accordingly,
with cryostatic temperatures of $k_{B}T\sim \SI{10}{\mu eV}$ being routinely
realized in the lab \cite{johnson05}, this condition can be met by
applying an external magnetic field of $\sim5-10 \SI{}{T}$ which is equivalent
to $\omega_{0}\approx100-200 \SI{}{\mu eV}$ in GaAs \cite{hanson07, hanson03}. Lastly,
the charging energy $U$, typically $\sim1-4 \SI{}{meV}$ \cite{vanderWiel02, ono02},
sets the largest energy scale in the problem justifying the
Coulomb-blockade regime with negligible double occupancy of the QD
provided that the chemical potential of the left lead is well below
the doubly occupied level. Lastly, we note that similar setups to the one
proposed here have previously been realized experimentally by Hanson et al.\cite{hanson04, hanson03}. 

Proceeding from these considerations, we will now show by numerical
simulation that an SR peaking of several orders of magnitude can be
observed for experimentally relevant parameters in the leakage current
through a quantum dot in the spin-blockade regime. We will first consider
the idealized case of homogeneous coupling for which an exact numerical
treatment is feasible even for a larger number of coupled nuclei.
Then, we will continue with the more realistic case of inhomogeneous
coupling for which an approximative scheme is applied. 
Here, we will also study scenarios in which the nuclear spins are 
not fully polarized initially. Finally, we will discuss intrinsic nuclear
dephasing effects and undesired cotunneling processes which have been 
omitted in our simulations.

\subsection{Superradiant Emission of Electrons}

\subsubsection{Idealized Setting}

\begin{figure}
\begin{center}
\includegraphics[width=\linewidth]{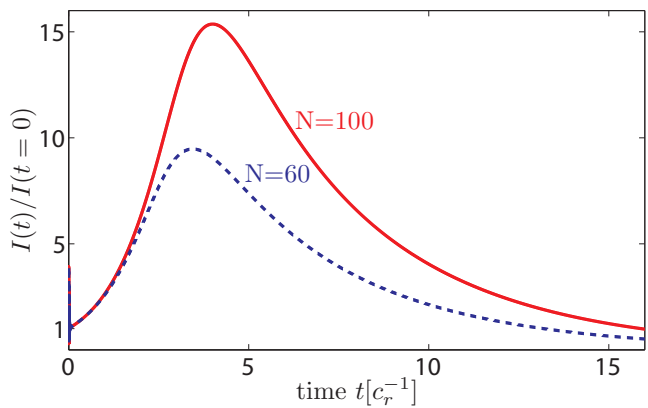}

\caption{\label{fig:time-evolution-homo}(color online). Typical time-evolution of the normalized
current for homogeneous coupling under dynamical compensation of the Overhauser field and a relative coupling strength of $\epsilon=0.5$, shown
here for $N=60$ and $N=100$ nuclear spins.   
The characteristic feature of superradiance, a pronounced peak in the leakage current proportional to $N$, is clearly observed. 
}
\end{center}
\end{figure}

\begin{figure}
\begin{center}
\includegraphics[width=1.1 \linewidth]{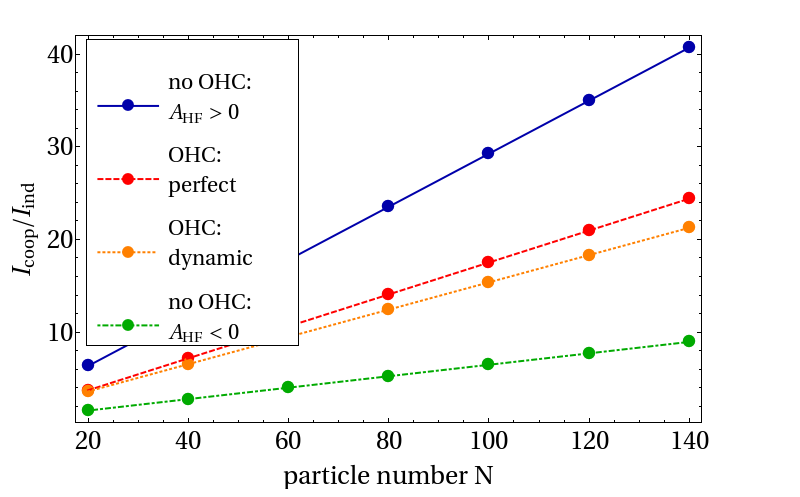}

\caption{\label{fig:SR-ratio-homo}(color online). Ratio of the maximum current to the initial
current $I_{\mathrm{coop}}/I_{\mathrm{ind}}$ as a function of the number of nuclear
spins $N$ for homogeneous coupling and a relative coupling strength of $\epsilon=0.5$: 
Results for perfect compensation (dashed line) are compared to the case
of dynamic compensation (dotted line) of the Overhauser field (OHC). Simulations without compensation of the Overhauser field set
bounds for the enhancement of the leakage current, depending on the sign of the HF coupling
constant $A_{\mathrm{HF}}$; solid and dash-dotted line for $A_{\mathrm{HF}}>0$ and $A_{\mathrm{HF}}<0$, respectively.  
}
\end{center}
\end{figure}

The homogeneous case allows for an exact treatment even for a relatively
large number of nuclei as the system evolves within the totally symmetric
low-dimensional subspace $\left\{ \left|J,m\right\rangle ,m=-J,\dots,J\right\} $.
Starting from a fully-polarized state, a strong intensity enhancement is
observed; typical results obtained from numerical simulations of Eqn.\eqref{eq:full-QME-OHC}
are depicted in Fig.~\ref{fig:time-evolution-homo} for $N=60$ and
$N=100$ nuclear spins. The corresponding relative peak heights display
a linear dependence with $N$, cf. Fig.~\ref{fig:SR-ratio-homo},
which we identify as the characteristic feature of superradiance. 
Here, we have used the numerical parameters 
$A_{\mathrm{HF}}=1$, $\omega_{0}=1$ and $\alpha=\beta_{\uparrow}^{\left(L\right)}=\beta_{\downarrow}^{\left(L\right)}=\beta_{\downarrow}^{\left(R\right)}=0.1$
in units of $\sim \SI{100} {\mu eV}$,
corresponding to a relative coupling strength $\epsilon=0.5$.

Before we proceed, some further remarks on the dynamic compensation
of the Overhauser field seem appropriate: We have merely introduced
it in our analysis in order to provide a clear criterion for the
presence of purely collective effects, given by $I_{\mathrm{coop}}/I_{\mathrm{ind}}>1$.
In other words, dynamic compensation of the Overhauser field is not
a necessary requirement for the observation of collective effects,
but it is rather an adequate tool to display them clearly. From an
experimental point of view, the dynamic compensation of the Overhauser
field might be challenging as it requires accurate knowledge about
the evolution of the nuclear spins. Therefore, we also present results
for the case in which the external magnetic field is constant and
no compensation is applied. This markedly relaxes the experimental
challenges. Here, we can distinguish two cases: Depending on the sign
of the HF coupling constant $A_{\mathrm{HF}}$, the time-dependence of the effective
Zeeman-splitting $\omega$ can either give rise to an additional enhancement
of the leakage current $\left(A_{\mathrm{HF}}>0\right)$ or it can counteract the
collective effects $\left(A_{\mathrm{HF}}<0\right)$. As shown in Fig.~\ref{fig:SR-ratio-homo},
this sets lower and upper bounds for the observed enhancement of the
leakage current. 

In Fig.~\ref{fig:SR-ratio-homo} we also compare the results obtained for dynamic compensation of
the Overhauser field to the idealized case of perfect compensation
in which the effect of the Overhauser field is set to zero, i.e., $H_{\mathrm{OH}}=gA^{z}S^{z}=0$.
Both approaches display the same features justifying our approximation
of neglecting residual (de)tuning effects of the dynamically compensated Overhauser
field w.r.t. the external Zeeman splitting $\omega_{0}$. 
This will also be discussed in greater detail below. 

\subsubsection{Beyond the Idealized Setting}

We now turn to the more realistic case of inhomogeneous coupling which in principle
could prevent the phasing necessary for SR. However, as shown
below, SR is still present in realistic inhomogeneous systems. In
contrast to the idealized case of homogeneous coupling, the dynamics
cannot be restricted to a low-dimensional subspace so that an exact
numerical treatment is not feasible due to the large number of nuclei.
We therefore use an approximate approach which has previously been
shown to capture the effect of nuclear spin coherences while allowing
for a numerical treatment of hundreds of spins \cite{christ07, kessler10}.
For simplicity, we restrict ourselves to the local moment regime in
which the current can be obtained directly from the electron inversion
$I\left(t\right)\propto\left\langle S^{+}S^{-}\right\rangle _{t}$.
By Eqn.\eqref{eq:QME-local-moment}, this expectation value is related
to a hierachy of correlation terms involving both the electron and
nuclear spins. Based on a Wick type factorization scheme, higher order
expressions are factorized in terms of the covariance matrix $\gamma_{ij}^{+}=\left\langle \sigma_{i}^{+}\sigma_{j}^{-}\right\rangle $
and the {}``mediated covariance matrix'' $\gamma_{ij}^{-}=\left\langle \sigma_{i}^{+}S^{z}\sigma_{j}^{-}\right\rangle $.
For further details, see Refs. \cite{christ07, kessler10}.
Moreover, the coupling constants $g_{j}$ have been obtained from 
the assumption of a two-dimensional Gaussian spatial electron
wavefunction of width $\sqrt{N}/2$.
Specifically, we will present results for two sets of numerical parameters, 
corresponding to a relative coupling strength of $\epsilon=0.5$, where 
$A_{\mathrm{HF}}=1$, $\omega_{0}=1$, $\gamma=0.1$ and $\Gamma=0.08$,  
and $\epsilon=0.55$ with
$A_{\mathrm{HF}}=1$, $\omega_{0}=0.9$, $\gamma=0.1$ and $\Gamma=0.067$.

\begin{figure}
\begin{center}
\includegraphics[width=0.95\linewidth]{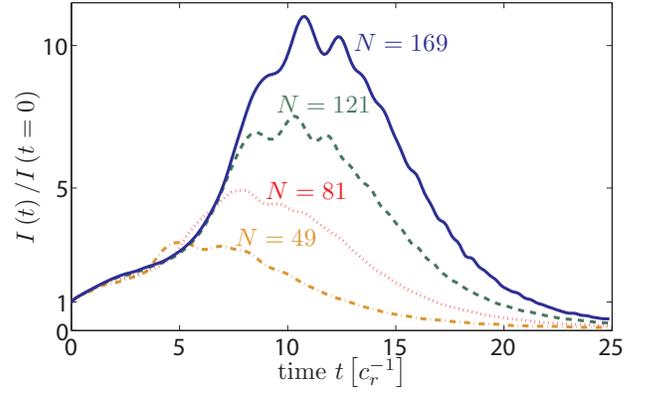}

\caption{\label{fig:time-evolution-inhomo}(color online). Typical time-evolution of the normalized
current for inhomogeneous coupling, shown here for up to $N=13^{2}$
nuclear spins and a relative coupling strength $\epsilon=0.55$. Compared to the idealized case of homogeneous coupling, the SR effects
are reduced, but still clearly present.  
A Gaussian spatial electron wave function has been
assumed and the Overhauser field is compensated dynamically. 
}
\end{center}
\end{figure}

\begin{figure}
\begin{center}
\includegraphics[width=\linewidth]{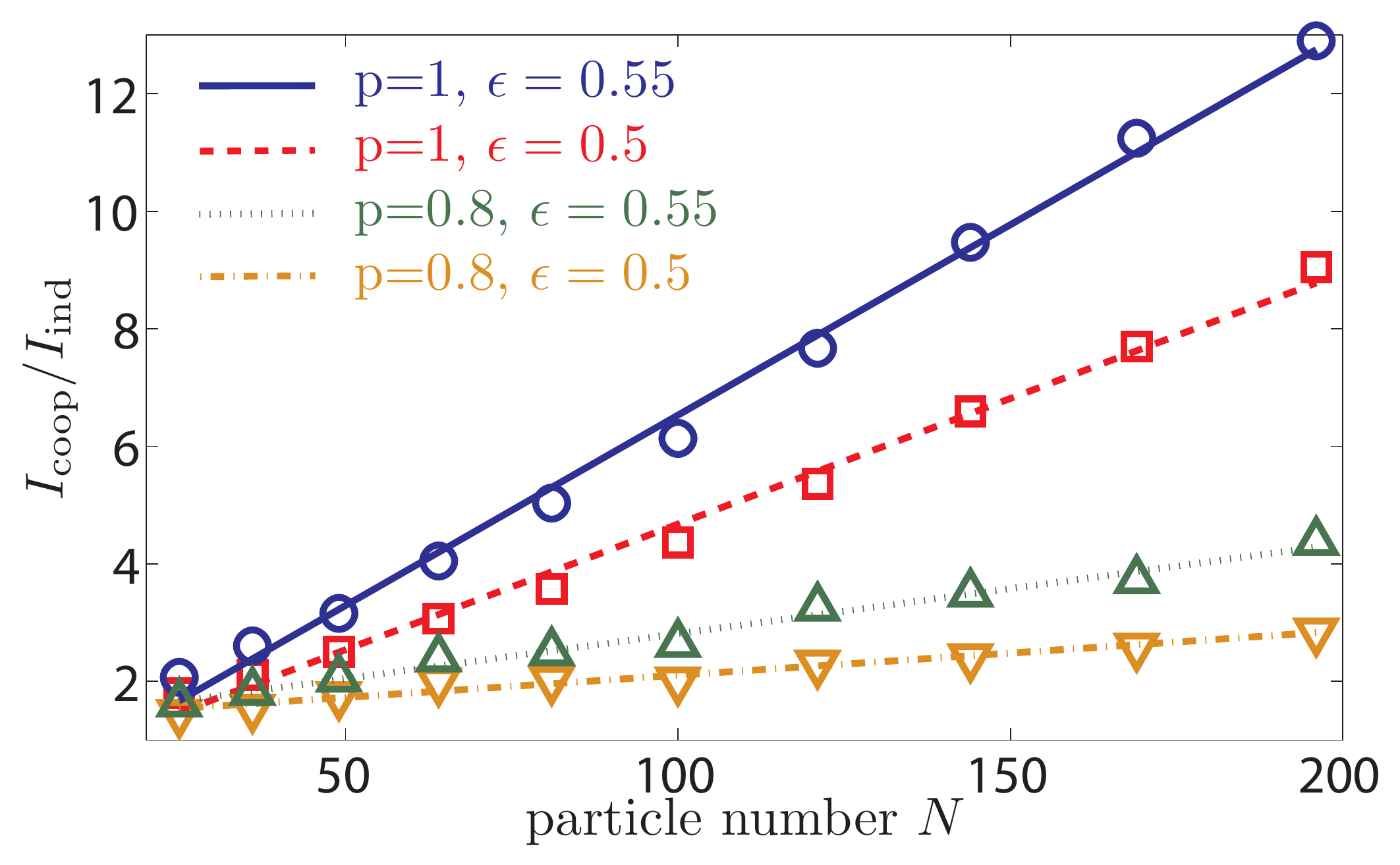}

\caption{\label{fig:SR-linear-inhomo}(color online). Ratio of the maximum current to the initial
current $I_{\mathrm{coop}}/I_{\mathrm{ind}}$ as a function of the number of nuclear
spins $N$ for relative coupling strengths $\epsilon=0.5$ and $\epsilon=0.55$: 
Results for inhomogenous coupling. The linear dependence
is still present when starting from a nuclear state with finite polarization
$p=0.8$. 
}
\end{center}
\end{figure}

As shown in Fig.~\ref{fig:time-evolution-inhomo} and \ref{fig:SR-linear-inhomo},
the results obtained with these methods demonstrate clear SR signatures.
In comparison to the ideal case of homogeneous coupling, the relative
height is reduced, but for a fully polarized initial state 
we still find a linear enhancement $I_{\mathrm{coop}}/I_{\mathrm{ind}}\approx0.043N\,\left(\epsilon=0.5\right)$; 
therefore, as long as this linear dependence is valid, for typically
$N\approx10^{5}-10^{6}$ a strong intensity enhancement of several
orders of magnitude is predicted $\left(\sim10^{3}-10^{4}\right)$. 

If the initial state is not fully polarized, SR effects are reduced:
However, when starting from a mixture of symmetric Dicke states $\left|J,J\right\rangle $
with polarization $p=80(60)\%$, we find that the linear $N$ dependence is still
present: $I_{\mathrm{coop}}/I_{\mathrm{ind}}\approx0.0075(0.0028)N\,\left(\epsilon=0.5\right)$,
i.e., the scaling is about a factor of $\sim5(15)$ weaker than for full polarization \cite{finitepol}.
Still, provided the linear scaling holds up to an experimentally realistic number of nuclei $N\approx10^{5}-10^{6}$,
this amounts to a relative enhancement of the order of $I_{\mathrm{coop}}/I_{\mathrm{ind}}\sim10^{2}-10^{3}$.

\begin{figure}
\begin{center}
\includegraphics[width=1.07\linewidth]{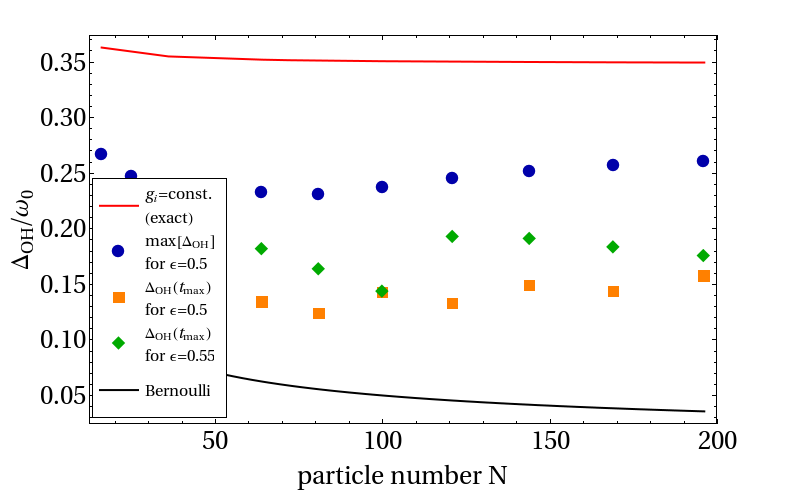}
\caption{\label{fig:deltaOH}(color online). Fluctuations of the Overhauser field
relative to the external Zeeman splitting $\omega_{0}$. In the limit of homogeneous HF coupling,
strong fluctuations build up towards the middle of the emission process (red line, $\epsilon=0.5$).
For inhomogeneous coupling this build-up of fluctuations is hindered by the dephasing between the 
nuclear spins, resulting in considerably smaller fluctuations: 
The value of the Overhauser fluctuations is shown at the time of the SR peak $t_{\mathrm{max}}$
for $\epsilon=0.5$ (orange squares) and $\epsilon=0.55$ (green diamonds). 
The Overhauser fluctuations reach a maximum value later than $t_{\mathrm{max}}$, see
blue dots for $\epsilon=0.5$. 
For independent homogeneously coupled nuclear spins, one can estimate the fluctuations via 
the Binominal distribution (black line). 
}
\end{center}
\end{figure}

In our simulations we have self-consistently verified that the fluctuations of
the Overhauser field, defined via 
\begin{equation}
\Delta_{\mathrm{OH}}\left(t\right)=g \sqrt{\left<A_{z}^{2}\right>_{t}-\left<A_{z}\right>^{2}_{t}},
\end{equation}
are indeed small compared to the external Zeeman splitting $\omega_{0}$ 
throughout the entire evolution. This ensures the validity of our perturbative approach 
and the realization of the spin-blockade regime.
From atomic superradiance it is known that in the limit of homogeneous coupling 
large fluctuations can build up, since in the middle of the emission process
the density matrix becomes a broad distribution over the Dicke states \cite{gross82}. 
Accordingly, in the idealized, exactly solvable case of homogeneous coupling 
we numerically find rather large fluctuations of the Overhauser field;
as demonstrated in Fig.~\ref{fig:deltaOH}, this holds independently of $N$.  
In particular, for a relative coupling strength $\epsilon=0.5$ the fluctuations culminate in
$\max\left[\Delta_{\mathrm{OH}}\right] / \omega_{0} \approx 0.35$.
However, in the case of inhomogeneneous HF coupling
the Overhauser field fluctuations are found to be smaller as
the build-up of these fluctuations is hindered by the Knight term causing 
dephasing among the nuclear spins.
As another limiting case, we also estimate the fluctuations 
for completely independent homogeneously coupled nuclear spins
via the Binominal distribution as $\max\left[\Delta_{\mathrm{OH}}\right] \sim 0.5 A_{\mathrm{HF}} / \sqrt{N}$
\cite{binomial}.
Moreover, we have also ensured self-consistently the validity of the perturbative 
treatment of the flip-flop dynamics; that is, throughout the entire evolution, even for maximum 
operative matrix elements $\left<A^{+}A^{-}\right>_{t}$, the strength of the flip-flop dynamics
$\| H_{\mathrm{ff}} \|$ was still at least five times smaller than $\omega_{0}$.  

\begin{figure}
\begin{center}
\includegraphics[width=1.0\linewidth]{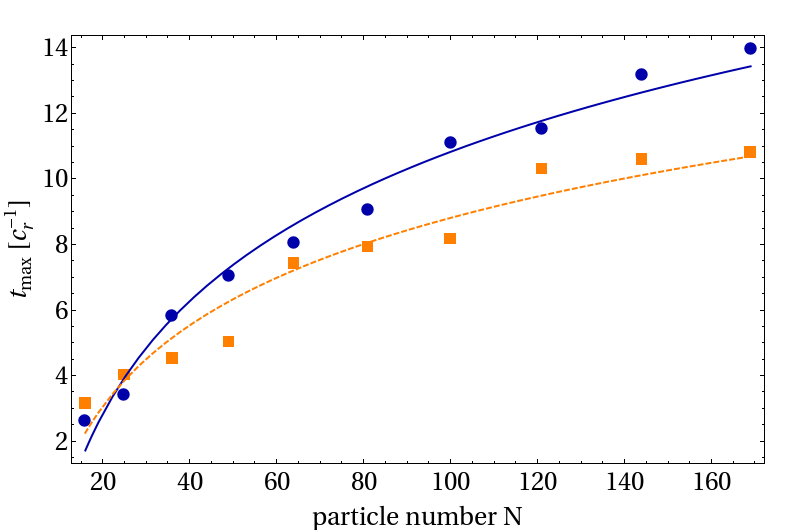}
\caption{\label{fig:duration}(color online). Total time till the observation of the
characteristic SR peaking $t_{\mathrm{max}}$ for $\epsilon=0.5$ (blue dots) and $\epsilon=0.55$ 
(orange squares). Based on Eqn.\eqref{eq:process-duration}, logarithmic fits are obtained from which we
estimate $t_{\mathrm{max}}$ for experimentally realistic number of nuclear spins $N\approx10^{5}$. 
}
\end{center}
\end{figure}

Initially, the HF mediated superradiance dynamics is rather slow, with its characteristic 
time scale set by $c_{r}^{-1}$; for experimentally realistic parameters -- in what follows we use
the parameter set   
$\left(\epsilon=0.5,\,\alpha\approx \SI{10}{\mu eV},\, N\approx10^{5}\right)$ for numerical estimates --
this corresponds to 
$c_{r}^{-1}\approx \SI{10}{\mu s}$. Based on fits as shown in Fig.~\ref{fig:duration}, we then 
estimate for the SR process duration $\left<t_{D}\right> \approx 50 c_{r}^{-1} \approx \SI{500}{\mu s}$
which is still smaller than recently reported \cite{takahashi11} nuclear 
decoherence times of $\sim \SI{1}{m s}$.
Therefore, it should be possible to observe the characteristic enhancement
of the leakage current before the nuclear spins decohere.   

Accordingly, in the initial phasing stage, the HF mediated lifting of the spin-blockade is rather weak
resulting in a low leakage current, approximatively given by
$I\left(t=0\right)\approx\epsilon^{2}\alpha/N$. Therefore, the initial
current due to HF processes is inversely proportional to the number
of nuclear spins $N$. However, as correlations among the nuclei build
up, the HF mediated lifting becomes more efficient culminating in
a maximum current of $I_{\mathrm{max}}\approx\epsilon^{2}\alpha$,
independent of $N$. For realistic experimental values --
also taking into account the effects of inhomogeneous HF coupling and 
finite initial polarization $p\approx0.6$ --
we estimate the initial (maximum) leakage current to be of the order
of $I\left(t=0\right)\approx \SI{6}{fA}\left(I_{\mathrm{max}}\approx \SI{10}{pA}\right)$.
Leakage currents in this range of magnitudes have already been detected
in single QD spin-filter experiments \cite{hanson04} as well as double
QD Pauli-blockade experiments \cite{koppens05, ono04, kobayashi11,ono02};
here, leakage currents below $\SI{10}{fA}$ and $\SI{150}{fA}$, respectively, have
been attributed explicitly to other spurious processes \cite{hanson04, kobayashi11}.
Among others, these will be addressed in greater detail in the following. 

In our simulations we have disregarded species inhomogeneities in the nuclear Zeeman energies.
In principle, these are large enough to cause additional dephasing
between the nuclear spins, similar to the inhomogeneous Knight field.
However, this dephasing mechanism only applies to nuclei of different
Zeeman energies, that is nuclei which belong to different species
\cite{christ07}. This leads to two or three mutually decohered subsystems
each of which is described by our theory. Moreover, we have neglected
the dipolar and quadrupolar interactions among the nuclear spins.
First, the latter is absent for nuclear spin $I=1/2$ (CdSe QDs) or
strain-free QDs \cite{bracker05}. Second, the nuclear dipole-dipole interaction can
cause diffusion and dephasing processes. Diffusion processes that
can change $A^{z}$ are strongly detuned and therefore of minor importance,
as corroborated by experimentally measured spin diffusion rates \cite{paget82, ota07}.
Resonant processes such as $\propto I_{i}^{z}I_{j}^{z}$ can lead
to dephasing similar to the inhomogenous Knight shift. This competes
with the phasing necessary for the observation of SR as expressed
by the first term in Eqn.\eqref{eq:QME-nuclear-spins}. The SR process
is the weakest at the very beginning of the evolution where we estimate
its strength as $c_{r}^{\mathrm{min}}\approx \SI{10}{\mu eV}/N$. An upper
bound for the dipole-dipole interaction in GaAs has been given in
Ref. \cite{schliemann03} as $\sim \SI{d-5}{\mu eV}$, in agreement with values
given in Refs. \cite{taylor07, bluhm10}. Therefore, the nuclear dipole-dipole
interaction can safely be neglected for $N\lesssim10^{5}$. In particular,
its effect should be further reduced for highly polarized ensembles.
Moreover, as argued above due to the presence of the MPM-term in Eqn.\eqref{eq:QME-nuclear-spins}
and demonstrated by our simulations, the observation of SR is even robust
against dephasing caused by the much stronger Knight field. 

Our transport setting is tuned into the sequential tunneling regime
and therefore we have disregarded cotunneling processes which are
fourth order in $H_{T}$. In principle, cotunneling processes could
lift the spin-blockade and add an extra contribution to the leakage
current that is independent of the HF dynamics. However, note that
cotunneling current scales as $I_{\mathrm{ct}}\propto\alpha^{2}$,
whereas sequential tunneling current $I\propto\alpha$; accordingly,
cotunneling current can always be suppressed by making the tunnel
barriers less transparent \cite{hanson04}. Moreover, inelastic cotunneling
processes exciting the QD spin can be ruled out for $eV,k_{B}T<\omega_{0}$
due to energy conservation \cite{recher00}. The effectiveness of a single quantum
dot to act as an electrically tunable spin filter has also been demonstrated
experimentally \cite{hanson04}: The spin-filter efficiency was measured
to be nearly $100\%$, with $I_{\mathrm{ct}}$ being smaller than
the noise floor $\sim \SI{10}{fA}$. Its actual value has been calculated
as $\sim \SI{d-4}{fA}$, from which we roughly estimate $I_{\mathrm{ct}}\sim \SI{d-2}{fA}$
in our setting. This is smaller than the initial HF mediated current
$I\left(t=0\right)$ and considerably smaller than $I_{\mathrm{max}}$,
even for an initially not fully polarized nuclear spin ensemble. Still,
if one is to explore the regime where cotunneling cannot be neglected,
phenomenological dissipative terms -- effectively describing the corresponding
spin-flip and pure dephasing mechanisms for inelastic and elastic
processes respectively -- should be added to Eqn.\eqref{eq:full-QME-OHC}.





\section{Conclusion and Outlook \label{sec:Conclusion}}

In summary, we have developed a master equation based theoretical
framework for nuclear spin assisted transport through a QD.
Due to the collective nature of the HF interaction, it incorporates 
intriguing many-body effects as well as
feedback mechanisms between the electron spin and nuclear spin dynamics.
As a prominent application, we have shown that the current through
a single electrically defined QD in the spin-blockade regime naturally exhibits
superradiant behavior. This effect stems from the collective hyperfine
interaction between the QD electron and the nuclear spin ensemble
in the QD. Its most striking feature is a lifting of the spin-blockade
and a sudden peak in the leakage current. The experimental observation
of this effect would provide clear evidence of coherent HF dynamics
of nuclear spin ensembles in QDs. 

Finally, we highlight possible directions of research going beyond 
our present work:
Apart from electronic superradiance, 
the setup proposed here is inherently well suited for other experimental 
applications like dynamic polarization of nuclear spins (DNP): 
In analogy to optical pumping, Eqn.\eqref{eq:QME-nuclear-spins} describes
\textit{electronic} pumping of the nuclear spins. Its steady states
are eigenstates of $A^{z}$, which lie in the kernel of the collective
jump-operator $A^{-}$. In particular, for a completely inhomogeneous system
the only steady state is the fully polarized one, the ideal initial state required for
the observation of SR effects. 
When starting from a completely unpolarized nuclear state, the uni-directionality of
Eqn.\eqref{eq:QME-nuclear-spins} 
-- electrons with one spin orientation exchange excitations with the nuclear spins,
while electrons of opposite spin primarily do not --
implies that the rather warm electronic reservoir 
can still extract entropy out of the nuclear system.  
More generally, the transport setting
studied here possibly opens up the route towards the (feedback-based)
electronic preparation of particular nuclear states in single QDs.
This is in line with similar ideas previously developed in double QD settings,
see e.g. Refs.\cite{baugh07, koppens05, kobayashi11, rudner07, takahashi11}.

In this work we have specialized on a single QD. 
However, our theory could be extended to a double QD (DQD) setting which is likely
to offer even more possibilities. DQDs are routinely operated in the 
Pauli-blockade regime where despite the presence of an applied source-drain voltage
the current through the device is blocked whenever the electron tunneling into
the DQD has the same spin orientation as the one already present.
The DQD parameters and the external magnetic field can be tuned such that 
the role of the states $\left|\sigma \right>, \sigma=\downarrow,\uparrow$, in our model  
is played by a pair of singlet and triplet states, while all other states are off-resonant.
Then, along the lines of our study, non-linearities appear due to dependencies 
between the electronic and nuclear subsystems and collective effects enter via the HF-mediated 
lifting of the spin-blockade.  

While we have focused on the Markovian regime and the precise conditions
for the validity thereof, Eqn.\eqref{eq:QME-non-Markovian-superoperator}
offers a starting point for studies of non-Markovian effects in the 
proposed transport setting. 
All terms appearing in the memory kernel of Eqn.\eqref{eq:QME-non-Markovian-superoperator}
are quadratic in the fermionic creation and annihilation operators 
allowing for an efficient numerical simulation, 
without having to explicitly invoke the flatness of the spectral density 
of the leads.
This should then shed light on possibly abrupt changes in the QD transport 
properties due to feedback mechanism between the nuclear spin ensemble 
and the electron spin.   


Lastly, our work also opens the door towards studies of dissipative phase
transitions in the transport setting: when combined with driving, the SR
dynamics can lead to a variety of strong-correlation effects,
non-equilibrium and dissipative phase transitions
\cite{brandes04,carmichael80,morrison08,kessler12}, which could
now be studied in a mesoscopic solid state system, complementing other
approaches to dissipative phase transitions in QDs \cite{chung09,borda06,legett87,rudner10}.

\begin{acknowledgments}
We acknowledge support by the DFG within SFB 631
and the Cluster of Excellence NIM. 
\end{acknowledgments}

\appendix

\begin{widetext}

\section{Microscopic Derivation of the Master Equation \label{sec:Derivation-QME}}

In this Appendix we provide some details regarding the derivation
of the master equations as stated in Eqn.\eqref{eq:QME} and
Eqn.\eqref{eq:full-QME-OHC}. It comprises the effect of
the HF dynamics in the memory kernel of Eqn.\eqref{eq:nakajima-zwanzig-after-Born}
and the subsequent approximation of independent rates of variation. 


In the following, we will show that it is self-consistent to neglect
the effect of the HF dynamics $\mathcal{L}_{1}\left(t\right)$ in
the memory-kernel of Eqn.\eqref{eq:nakajima-zwanzig-after-Born} provided
that the bath correlation time $\tau_{c}$ is short compared to the
Rabi flips produced by the HF dynamics. This needs to be addressed
as cooperative effects potentially drive the system from a weakly
coupled into a strongly coupled regime. First, we reiterate the Schwinger-Dyson
identity in Eqn.\eqref{eq:Schwinger-Dyson} as an infinite sum over
time-ordered nested commutators \begin{equation}
e^{-i\left(\mathcal{L}_{0}+\mathcal{L}_{1}\right)\tau}=e^{-i\mathcal{L}_{0}\tau}\sum_{n=0}^{\infty}\left(-i\right)^{n}\int_{0}^{\tau}d\tau_{1}\int_{0}^{\tau_{1}}d\tau_{2}\dots\int_{0}^{\tau_{n-1}}d\tau_{n}\,\tilde{\mathcal{L}}_{1}\left(\tau_{1}\right)\tilde{\mathcal{L}}_{1}\left(\tau_{2}\right)\dots\tilde{\mathcal{L}}_{1}\left(\tau_{n}\right),\label{eq:time-ordered-Dyson}\end{equation}
where for any operator $X$\begin{equation}
\tilde{\mathcal{L}}_{1}\left(\tau\right)X=e^{i\mathcal{L}_{0}\tau}\mathcal{L}_{1}e^{-i\mathcal{L}_{0}\tau}X=\left[e^{iH_{0}\tau}H_{1}e^{-iH_{0}\tau},X\right]=\left[\tilde{H}_{1}\left(\tau\right),X\right].\end{equation}
More explicitly, up to second order Eqn.\eqref{eq:time-ordered-Dyson}
is equivalent to \begin{eqnarray}
e^{-i\left(\mathcal{L}_{0}+\mathcal{L}_{1}\right)\tau}X & = & e^{-i\mathcal{L}_{0}\tau}X-ie^{-i\mathcal{L}_{0}\tau}\int_{0}^{\tau}d\tau_{1}\left[\tilde{H}_{1}\left(\tau_{1}\right),X\right]\nonumber \\
 &  & -e^{-i\mathcal{L}_{0}\tau}\int_{0}^{\tau}d\tau_{1}\int_{0}^{\tau_{1}}d\tau_{2}\left[\tilde{H}_{1}\left(\tau_{1}\right),\left[\tilde{H}_{1}\left(\tau_{2}\right),X\right]\right]+\dots\end{eqnarray}
Note that the time-dependence of $\tilde{H}_{1}\left(\tau\right)$
is simply given by \begin{equation}
\tilde{H}_{1}\left(\tau\right)=e^{i\omega\tau}H_{+}+e^{-i\omega\tau}H_{-}+H_{\Delta\mathrm{OH}},\,\,\,\,\,\,\,\,\,\, H_{\pm}=\frac{g}{2}S^{\pm}A^{\mp},\end{equation}
where the effective Zeeman splitting $\omega=\omega_{0}+g\left\langle A^{z}\right\rangle _{t}$
is time-dependent. Accordingly, we define $\tilde{\mathcal{L}}_{1}\left(\tau\right)=\tilde{\mathcal{L}}_{+}\left(\tau\right)+\tilde{\mathcal{L}}_{-}\left(\tau\right)+\tilde{\mathcal{L}}_{\Delta\mathrm{OH}}\left(\tau\right)=e^{i\omega\tau}\mathcal{L}_{+}+e^{-i\omega\tau}\mathcal{L}_{-}+\mathcal{L}_{\Delta\mathrm{OH}},$
where $\mathcal{L}_{x}\cdot=\left[H_{x},\cdot\right]$ for $x=\pm,\Delta\mathrm{OH}$.
In the next steps, we will explicitly evaluate the first two contributions
to the memory kernel that go beyond $n=0$ and then generalize our
findings to any order $n$ of the Schwinger-Dyson series.

\subsubsection*{First order correction}

The first order contribution $n=1$ in Eqn.\eqref{eq:nakajima-zwanzig-after-Born}
is given by\begin{equation}
\Xi^{\left(1\right)}=i\int_{0}^{t}d\tau\int_{0}^{\tau}d\tau_{1}\mathsf{Tr_{B}}\left(\mathcal{L}_{T}e^{-i\mathcal{L}_{0}\tau}\left[\tilde{H}_{1}\left(\tau_{1}\right),X\right]\right).\end{equation}
Performing the integration in $\tau_{1}$ leads to \begin{eqnarray}
\Xi^{\left(1\right)} & = & \int_{0}^{t}d\tau\left\{ \frac{g}{2\omega}\left(1-e^{-i\omega\tau}\right)\mathsf{Tr_{B}}\left(\mathcal{L}_{T}\left[S^{+}A^{-},\tilde{X}_{\tau}\right]\right)\right.\nonumber \\
 &  & +\frac{g}{2\omega}\left(e^{i\omega\tau}-1\right)\mathsf{Tr_{B}}\left(\mathcal{L}_{T}\left[S^{-}A^{+},\tilde{X}_{\tau}\right]\right)\nonumber \\
 &  & \left.+ig\tau\mathsf{Tr_{B}}\left(\mathcal{L}_{T}\left[\left(A^{z}-\left\langle A^{z}\right\rangle _{t}\right)S^{z},\tilde{X}_{\tau}\right]\right)\right\} \end{eqnarray}
where, for notational convenience, we introduced the operators $X=\mathcal{L}_{T}\rho_{S}\left(t-\tau\right)\rho_{B}^{0}$
and $\tilde{X}_{\tau}=e^{-iH_{0}\tau}\left[H_{T},\rho_{S}\left(t-\tau\right)\rho_{B}^{0}\right]e^{iH_{0}\tau}\approx\left[\tilde{H}_{T}\left(\tau\right),\rho_{S}\left(t\right)\rho_{B}^{0}\right]$.
In accordance with previous approximations, we have replaced $e^{-iH_{0}\tau}\rho_{S}\left(t-\tau\right)e^{iH_{0}\tau}$
by $\rho_{S}\left(t\right)$ since any additional term besides $H_{0}$
would be of higher order in perturbation theory \cite{timm08, harbola06}.
In particular, this disregards dissipative effects: In our case, this
approximation is valid self-consistently provided that the tunneling
rates are small compared to effective Zeeman splitting $\omega$.
The integrand decays on the leads-correlation timescale $\tau_{c}$
which is typically much faster than the timescale set by the effective
Zeeman splitting, $\omega\tau_{c}\ll1$. This separation of timescales
allows for an expansion in the small parameter $\omega\tau$, e.g.
$\frac{g}{\omega}\left(e^{i\omega\tau}-1\right)\approx ig\tau$. We
see that the first order correction can be neglected if the the bath
correlation time $\tau_{c}$ is sufficiently short compared to the
timescale of the HF dynamics, that is $g\tau_{c}\ll1$. The latter
is bounded by the total hyperfine coupling constant $A_{\mathrm{HF}}$ (since $\left|\left|gA^{x}\right|\right|\leq A_{\mathrm{HF}}$)
so that the requirement for disregarding the first order term reads
$A_{\mathrm{HF}}\tau_{c}\ll1$.

\subsubsection*{Second order correction}

The contribution of the second term $n=2$ in the Schwinger-Dyson
expansion can be decomposed into \begin{equation}
\Xi^{\left(2\right)}=\Xi_{\mathrm{zz}}^{\left(2\right)}+\Xi_{\mathrm{ff}}^{\left(2\right)}+\Xi_{\mathrm{fz}}^{\left(2\right)}.\end{equation}
The first term $\Xi_{\mathrm{zz}}^{\left(2\right)}$ contains contributions
from $H_{\Delta\mathrm{OH}}$ only \begin{eqnarray}
\Xi_{\mathrm{zz}}^{\left(2\right)} & = & \int_{0}^{t}d\tau\int_{0}^{\tau}d\tau_{1}\int_{0}^{\tau_{1}}d\tau_{2}\mathsf{Tr_{B}}\left(\mathcal{L}_{T}e^{-i\mathcal{L}_{0}\tau}\left[\tilde{H}_{\Delta\mathrm{OH}}\left(\tau_{1}\right),\left[\tilde{H}_{\Delta\mathrm{OH}}\left(\tau_{2}\right),X\right]\right]\right)\\
 & = & -\int_{0}^{t}d\tau\left(g\tau\right)^{2}\mathsf{Tr_{B}}\left[\mathcal{L}_{T}\left(\delta A^{z}S^{z}\tilde{X}_{\tau}\delta A^{z}S^{z}-\frac{1}{2}\left\{ \delta A^{z}S^{z}\delta A^{z}S^{z},\tilde{X}_{\tau}\right\} \right)\right]\end{eqnarray}
Similarly, $\Xi_{\mathrm{ff}}^{\left(2\right)}$ which comprises contributions
from $H_{\mathrm{ff}}$ only is found to be \begin{eqnarray}
\Xi_{\mathrm{ff}}^{\left(2\right)} & = & \frac{g^{2}}{4\omega^{2}}\int_{0}^{t}d\tau\left\{ \left(1+i\omega\tau-e^{i\omega\tau}\right)\mathsf{Tr_{B}}\left[\mathcal{L}_{T}\left(S^{+}S^{-}A^{-}A^{+}\tilde{X}_{\tau}+\tilde{X}_{\tau}S^{-}S^{+}A^{+}A^{-}\right)\right]\right.\nonumber \\
 &  & \left.+\left(1-i\omega\tau-e^{-i\omega\tau}\right)\mathsf{Tr_{B}}\left[\mathcal{L}_{T}\left(S^{-}S^{+}A^{+}A^{-}\tilde{X}_{\tau}+\tilde{X}_{\tau}S^{+}S^{-}A^{-}A^{+}\right)\right]\right\} .\end{eqnarray}
Here, we have used the following simplification: The time-ordered
products which include flip-flop terms only can be simplified to two
possible sequences in which $\mathcal{L}_{+}$ is followed by $\mathcal{L}_{-}$
and vice versa. This holds since \begin{equation}
\mathcal{L}_{\pm}\mathcal{L}_{\pm}X=\left[H_{\pm},\left[H_{\pm},X\right]\right]=H_{\pm}H_{\pm}X+XH_{\pm}H_{\pm}-2H_{\pm}XH_{\pm}=0.\end{equation}
Here, the first two terms drop out immediately since the electronic
jump-operators $S^{\pm}$ fulfill the relation $S^{\pm}S^{\pm}=0$.
In the problem at hand, also the last term gives zero because of particle
number superselection rules: In Eqn.\eqref{eq:nakajima-zwanzig-after-Born}
the time-ordered product of superoperators acts on $X=\left[H_{T},\rho_{S}\left(t-\tau\right)\rho_{B}^{0}\right]$.
Thus, for the term $H_{\pm}XH_{\pm}$ to be nonzero, coherences in
Fock space would be required which are consistently neglected; compare
Ref. \cite{harbola06}.
This is equivalent to ignoring coherences between the system and the
leads. Note that the same argument holds for any combination $H_{\mu}XH_{\nu}$
with $\mu,\nu=\pm$. 

Similar results can be obtained for $\Xi_{\mathrm{fz}}^{\left(2\right)}$
which comprises $H_{\pm}$ as well as $H_{\Delta\mathrm{OH}}$ in
all possible orderings. Again, using that the integrand decays on
a timescale $\tau_{c}$ and expanding in the small parameter $\omega\tau$
shows that the second order contribution scales as $\sim\left(g\tau_{c}\right)^{2}$.
Our findings for the first and second order correction suggest that
the $n$-th order correction scales as $\sim\left(g\tau_{c}\right)^{n}$.
This will be proven in the following by induction.

\subsubsection*{n-th order correction}

The scaling of the $n$-th term in the Dyson series is governed by
the quantities of the form 
\begin{equation}
\xi_{+-\dots}^{\left(n\right)}\left(\tau\right)=g^{n}\int_{0}^{\tau}d\tau_{1}\int_{0}^{\tau_{1}}d\tau_{2}\dots\int_{0}^{\tau_{n-1}}d\tau_{n}e^{i\omega\tau_{1}}e^{-i\omega\tau_{2}}\dots,
\end{equation}
where the index suggests the order in which $H_{\pm}$ (giving an
exponential factor) and $H_{\Delta\mathrm{OH}}$ (resulting in a factor
of 1) appear. Led by our findings for $n=1,2$, we claim that the
expansion of $\xi_{+-\dots}^{\left(n\right)}\left(\tau\right)$ for
small $\omega\tau$ scales as $\xi_{+-\dots}^{\left(n\right)}\left(\tau\right)\sim\left(g\tau\right)^{n}$.
Then, the $\left(n+1\right)$-th terms scale as 
\begin{eqnarray}
\xi_{-(\Delta\mathrm{OH})+-\dots}^{\left(n+1\right)}\left(\tau\right) & = & g^{n+1}\int_{0}^{\tau}d\tau_{1}\int_{0}^{\tau_{1}}d\tau_{2}\dots\int_{0}^{\tau_{n-1}}d\tau_{n}\int_{0}^{\tau_{n}}d\tau_{n+1}\left(\begin{array}{c}
e^{-i\omega\tau_{1}}\\
1\end{array}\right)e^{+i\omega\tau_{2}}\dots\\
 & = & g\int_{0}^{\tau}d\tau_{1}\left(\begin{array}{c}
e^{-i\omega\tau_{1}}\\
1\end{array}\right)\xi_{+-\dots}^{\left(n\right)}\left(\tau_{1}\right)\\
 & \sim & \left(g\tau\right)^{n+1}.
\end{eqnarray}
Since we have already verified this result for $n=1,2$, the general
result follows by induction. This completes the proof.

\section{Adiabatic Elimination of the QD Electron \label{sec:Adiabatic-Elimination}}

For a sufficiently small relative coupling strength $\epsilon$ the
nuclear dynamics are slow compared to the electronic QD dynamics.
This allows for an adiabatic elimination of the electronic degrees
of freedom yielding an effective master equation for the nuclear spins
of the QD. 

Our analysis starts out from Eqn.\eqref{eq:QME-local-moment} which
we write as 
\begin{equation}
\dot{\rho}=\mathcal{W}_{0}\rho+\mathcal{W}_{1}\rho,
\end{equation}
where 
\begin{eqnarray}
\mathcal{W}_{0}\rho & = & -i\left[\omega_{0}S^{z},\rho\right]+\gamma\left[S^{-}\rho S^{+}-\frac{1}{2}\left\{ S^{+}S^{-},\rho\right\} \right]+\Gamma \left[S^{z} \rho S^{z}-\frac{1}{4}\rho\right]\\
\mathcal{W}_{1}\rho & = & -i\left[H_{\mathrm{HF}},\rho\right].
\end{eqnarray}
Note that the superoperator $\mathcal{W}_{0}$ only acts on the electronic
degrees of freedom. It describes an electron in an external magnetic
field that experiences a decay as well as a pure dephasing mechanism.
In zeroth order of the coupling parameter $\epsilon$ the electronic
and nuclear dynamics of the QD are decoupled and SR effects cannot
be expected. These are contained in the interaction term $\mathcal{W}_{1}$.

Formally, the adiabatic elimination of the electronic degrees of freedom
can be achieved as follows \cite{cirac92}: To zeroth order in $\epsilon$
the eigenvectors of $\mathcal{W}_{0}$ with zero eigenvector $\lambda_{0}=0$
are 
\begin{equation}
\mathcal{W}_{0}\mu\otimes\rho_{SS}=0,
\end{equation}
where $\rho_{SS}=\left|\downarrow\right\rangle \left\langle \downarrow\right|$
is the stationary solution for the electronic dynamics and $\mu$
describes some arbitrary state of the nuclear system. The zero-order
Liouville eigenstates corresponding to $\lambda_{0}=0$ are coupled
to the subspaces of {}``excited'' nonzero (complex) eigenvalues
$\lambda_{k}\neq0$ of $\mathcal{W}_{0}$ by the action of $\mathcal{W}_{1}$.
Physically, this corresponds to a coupling between electronic and
nuclear degrees of freedom. In the limit where the HF dynamics are
slow compared to the electronic frequencies, i.e. the Zeeman splitting
$\omega_{0}$, the decay rate $\gamma$ and the dephasing rate $\Gamma$,
the coupling between these blocks of eigenvalues and Liouville subspaces
of $\mathcal{W}_{0}$ is weak justifying a perturbative treatment.
This motivates the definition of a projection operator $P$ onto the
subspace with zero eigenvalue $\lambda_{0}=0$ of $\mathcal{W}_{0}$
according to 
\begin{equation}
P\rho=\mathsf{Tr_{el}}\left[\rho\right]\otimes\rho_{SS}=\mu\otimes\left|\downarrow\right\rangle \left\langle \downarrow\right|,
\end{equation}
where $\mu=\mathsf{Tr_{el}}\left[\rho\right]$ is a density operator
for the nuclear spins, $\mathsf{Tr_{el}}\dots$ denotes the trace
over the electronic subspace and by definition $\mathcal{W}_{0}\rho_{SS}=0$.
The complement of $P$ is $Q=1-P$. By projecting the master equation
on the $P$ subspace and tracing over the electronic degrees of freedom
we obtain an effective master equation for the nuclear spins in second
order perturbation theory
\begin{equation}
\dot{\mu}=\mathsf{Tr_{el}}\left[P\mathcal{W}_{1}P\rho-P\mathcal{W}_{1}Q\mathcal{W}_{0}^{-1}Q\mathcal{W}_{1}P\rho\right].
\end{equation}
Using $\mathsf{Tr_{el}}\left[S^{z}\rho_{SS}\right]=-1/2$, the first
term is readily evaluated and yields the Knight shift seen by the
nuclear spins 
\begin{equation}
\mathsf{Tr_{el}}\left[P\mathcal{W}_{1}P\rho\right]=+i\frac{g}{2}\left[A^{z},\mu\right].\label{eq:elimination-Knight}
\end{equation}
The derivation of the second term is more involved. It can be rewritten
as 
\begin{eqnarray}
-\mathsf{Tr_{el}}\left[P\mathcal{W}_{1}Q\mathcal{W}_{0}^{-1}Q\mathcal{W}_{1}P\rho\right] & = & -\mathsf{Tr_{el}}\left[P\mathcal{W}_{1}\left(1-P\right)\mathcal{W}_{0}^{-1}\left(1-P\right)\mathcal{W}_{1}P\rho\right]\\
 & = & \int_{0}^{\infty}d\tau\,\mathsf{Tr_{el}}\left[P\mathcal{W}_{1}e^{\mathcal{W}_{0}\tau}\mathcal{W}_{1}P\rho\right]-\int_{0}^{\infty}d\tau\,\mathsf{Tr_{el}}\left[P\mathcal{W}_{1}P\mathcal{W}_{1}P\rho\right].\label{eq:after-Laplace-trafo}
\end{eqnarray}
Here, we used the Laplace transform $-\mathcal{W}_{0}^{-1}=\int_{0}^{\infty}d\tau\, e^{\mathcal{W}_{0}\tau}$
and the property $e^{\mathcal{W}_{0}\tau}P=Pe^{\mathcal{W}_{0}\tau}=P$. 

Let us first focus on the first term in Eqn.\eqref{eq:after-Laplace-trafo}.
It contains terms of the form 
\begin{eqnarray}
\mathsf{Tr_{el}}\left[P\left[A^{+}S^{-},e^{\mathcal{W}_{0}\tau}\left[A^{-}S^{+},\mu\otimes\rho_{SS}\right]\right]\right] & = & \mathsf{Tr_{el}}\left[S^{-}e^{\mathcal{W}_{0}\tau}\left(S^{+}\rho_{SS}\right)\right]A^{+}A^{-}\mu\\
 &  & -\mathsf{Tr_{el}}\left[S^{-}e^{\mathcal{W}_{0}\tau}\left(S^{+}\rho_{SS}\right)\right]A^{-}\mu A^{+}\\
 &  & +\mathsf{Tr_{el}}\left[S^{-}e^{\mathcal{W}_{0}\tau}\left(\rho_{SS}S^{+}\right)\right]\mu A^{-}A^{+}\\
 &  & -\mathsf{Tr_{el}}\left[S^{-}e^{\mathcal{W}_{0}\tau}\left(\rho_{SS}S^{+}\right)\right]A^{+}\mu A^{-}
\end{eqnarray}
This can be simplified using the following relations: Since $\rho_{SS}=\left|\downarrow\right\rangle \left\langle \downarrow\right|$,
we have $S^{-}\rho_{SS}=0$ and $\rho_{SS}S^{+}=0$. Moreover, $\left|\uparrow\right\rangle \left\langle \downarrow\right|$
and $\left|\downarrow\right\rangle \left\langle \uparrow\right|$
are eigenvectors of $\mathcal{W}_{0}$ with eigenvalue $-\left(i\omega_{0}+\alpha/2\right)$
and $+\left(i\omega_{0}+\alpha/2\right)$, where $\alpha=\gamma+\Gamma$,
yielding 
\begin{eqnarray}
e^{\mathcal{W}_{0}\tau}\left(S^{+}\rho_{SS}\right) & = & e^{-\left(i\omega_{0}+\alpha/2\right)\tau}\left|\uparrow\right\rangle \left\langle \downarrow\right|\\
e^{\mathcal{W}_{0}\tau}\left(\rho_{SS}S^{-}\right) & = & e^{+\left(i\omega_{0}+\alpha/2\right)\tau}\left|\downarrow\right\rangle \left\langle \uparrow\right|.
\end{eqnarray}
This leads to 
\begin{equation}
\mathsf{Tr_{el}}\left[P\left[A^{+}S^{-},e^{\mathcal{W}_{0}\tau}\left[A^{-}S^{+},\mu\otimes\rho_{SS}\right]\right]\right]=e^{-\left(i\omega_{0}+\alpha/2\right)\tau}\left(A^{+}A^{-}\mu-A^{-}\mu A^{+}\right).\label{eq:elimination-1}
\end{equation}
Similarly, one finds 
\begin{equation}
\mathsf{Tr_{el}}\left[P\left[A^{-}S^{+},e^{\mathcal{W}_{0}\tau}\left[A^{+}S^{-},\mu\otimes\rho_{SS}\right]\right]\right]=e^{+\left(i\omega_{0}+\alpha/2\right)\tau}\left(\mu A^{+}A^{-}-A^{-}\mu A^{+}\right).\label{eq:elimination-2}
\end{equation}
Analogously, one can show that terms containing two flip or two flop
terms give zero. The same holds for mixed terms that comprise one
flip-flop and one Overhauser term with $\sim A^{z}S^{z}$. The term
consisting of two Overhauser contributions gives 
\begin{equation}
\mathsf{Tr_{el}}\left[P\left[A^{z}S^{z},e^{\mathcal{W}_{0}\tau}\left[A^{z}S^{z},\mu\otimes\rho_{SS}\right]\right]\right]=-\frac{1}{4}\left[2A^{z}\mu A^{z}-\left[A^{z}A^{z},\mu\right]\right].
\end{equation}
However, this term exactly cancels with the second term from Eqn.\eqref{eq:after-Laplace-trafo}.
Thus we are left with the contributions coming from Eqn.\eqref{eq:elimination-1}
and Eqn.\eqref{eq:elimination-2}. Restoring the prefactors of $-ig/2$,
we obtain 
\begin{eqnarray}
\mathsf{Tr_{el}}\left[P\mathcal{W}_{1}Q\left(-\mathcal{W}_{0}^{-1}\right)Q\mathcal{W}_{1}P\rho\right] & = & \frac{g^{2}}{4}\int_{0}^{\infty}d\tau\left[e^{-\left(i\omega_{0}+\alpha/2\right)\tau}\left(A^{-}\mu A^{+}-A^{+}A^{-}\mu\right)\right.\nonumber \\
 &  & \left.+e^{+\left(i\omega_{0}+\alpha/2\right)\tau}\left(A^{-}\mu A^{+}-\mu A^{+}A^{-}\right)\right].
\end{eqnarray}
Performing the integration and separating real from imaginary terms
yields
\begin{equation}
\mathsf{Tr_{el}}\left[P\mathcal{W}_{1}Q\left(-\mathcal{W}_{0}^{-1}\right)Q\mathcal{W}_{1}P\rho\right]=c_{r}\left[A^{-}\mu A^{+}-\frac{1}{2}\left\{ A^{+}A^{-},\mu\right\} \right]+ic_{i}\left[A^{+}A^{-},\mu\right],\label{eq:elimination-SR}
\end{equation}
where $c_{r}=g^{2}/\left(4\omega_{0}^{2}+\alpha^{2}\right)\alpha$
and $c_{i}=g^{2}/\left(4\omega_{0}^{2}+\alpha^{2}\right)\omega_{0}$.
Combining Eqn.\eqref{eq:elimination-Knight} with Eqn.\eqref{eq:elimination-SR}
directly gives the effective master equation for the nuclear spins
given in Eqn.\eqref{eq:QME-nuclear-spins} in the main text.

\end{widetext}


\begin{thebibliography}{50}
\bibitem{brandes04}T. Brandes. Coherent and collective quantum optical
effects in mesoscopic systems. Phys. Rep. \textbf{408}, 315 (2004).

\bibitem{awschalom02}D. D. Awschalom, N. Samarth, and D. Loss, 
\emph{Semiconductor Spintronics and Quantum Computation} (Springer-Verlag,
Berlin, 2002).

\bibitem{sharvin81}D. Y. Sharvin and Y. V. Sharvin. 
Magnetic-flux quantization in a cylindrical film of a normal metal.
JETP Lett. \textbf{34}, 272 (1981). 

\bibitem{vanWees88}B.J. van Wees, H. Van Houten, C.W.J. Beenacker,
J.G. Williamson, L.P. Kouwenhoven, D. van der Marel, and C.T. Foxon.
Quantized conductance of point contacts in a two-dimensional electron gas.
Phys. Rev. Lett. \textbf{60}, 848 (1988). 

\bibitem{wharam88}D. Wharam, T.J. Thornton, R. Newbury, M. Pepper,
H. Ahmed, J.E.F. Frost, D.G. Husko, D.C. Peacock, D.A. Ritchie, and G.A.C. Jones. 
One-dimensional transport and the quantisation of the ballistic resistance. 
J. Phys. C \textbf{21}, 209 (1988). 

\bibitem{nazarov09}Y. V. Nazarov and Y. M. Blanter, 
\emph{Quantum Transport} (Cambridge University Press, Cambridge, 2009).

\bibitem{datta97} S. Datta,  
\emph{Electronic Transport in Mesoscopic Systems},
(Cambridge University Press, Cambridge, 1997).

\bibitem{hanson07}R. Hanson, L. P. Kouwenhoven, J. R. Petta, S. Tarucha,
and L. M. Vandersypen. Spins in few-electron quantum dots. Rev. Mod.
Phys.\textbf{ 79}, 1217 (2007). 

\bibitem{vanderWiel02}W. G. van der Wiel, S. De Franceschi, J. M.
Elzerman, T. Fujisawa, S. Tarucha, and L. P. Kouwenhoven. Electron
transport through double quantum dots. Rev. Mod. Phys. \textbf{75},
1 (2002). 

\bibitem{johnson05}A. C. Johnson, J. R. Petta, J. M. Taylor, A. Yacoby,
M. D. Lukin, C. M. Marcus, M. P. Hanson, and A. C. Gossard. Triplet\textendash{}singlet
spin relaxation via nuclei in a double quantum dot. Nature \textbf{435},
925 (2005).

\bibitem{jouravlev06}O. N. Jouravlev and Y. V. Nazarov. Electron
transport in a double quantum dot governed by a nuclear magnetic field.
Phys. Rev. Lett. \textbf{96}, 176804 (2006). 

\bibitem{baugh07}J. Baugh, Y. Kitamura, K. Ono, and S. Tarucha. Large
nuclear Overhauser fields detected in vertically coupled quantum dots.
Phys. Rev. Lett. \textbf{99}, 096804 (2007). 

\bibitem{petta08}J. R. Petta, J. M. Taylor, A. C. Johnson, A. Yacoby,
M. D. Lukin, C. M. Marcus, M. P. Hanson, and A. C. Gossard. Dynamic
nuclear polarization with single electron spins. Phys. Rev. Lett.
\textbf{100}, 067601 (2008). 

\bibitem{inarrea07}J. I\~{n}arrea, G. Platero, and A. H. MacDonald. Electronic
transport through a double quantum dot in the spin-blockade regime:
Theoretical models. Phys. Rev. B \textbf{76}, 085329 (2007). 

\bibitem{koppens05}F. H. L. Koppens, J. A. Folk, J. M. Elzerman,
R. Hanson, L. H. Willems van Beveren, I. T. Vink, H. P. Tranitz, W.
Wegschneider, L. P. Kouwenhoven, and L. M. K. Vandersypen. Control
and detection of singlet-triplet mixing in a random nuclear field.
Science \textbf{309}, 1346 (2005). 

\bibitem{ono04}K. Ono and S. Tarucha. Nuclear-spin-induced oscillatory
current in spin-blockaded quantum dots. Phys. Rev. Lett. \textbf{92},
256803 (2004). 

\bibitem{pfund07}A. Pfund, I. Shorubalko, K. Ensslin, and R. Leturcq.
Suppression of spin relaxation in an InAs nanowire double quantum
dot. Phys. Rev. Lett. \textbf{99}, 036801 (2007).

\bibitem{kobayashi11}T. Kobayashi, K. Hitachi, S. Sasaki, and K.
Muraki. Observation of hysteretic transport due to dynamic nuclear
spin polarization in a GaAs lateral double quantum dot. Phys. Rev.
Lett. \textbf{107}, 216802 (2011).

\bibitem{ono02}K. Ono, D. G. Austing, Y. Tokura, and S. Tarucha.
Current rectification by Pauli exclusion in a weakly coupled double
quantum dot system. Science \textbf{297}, 1313 (2002).

\bibitem{rudner07}M. S. Rudner and L. S. Levitov. Self-polarization
and dynamical cooling of nuclear spins in double quantum dots. Phys.
Rev. Lett. \textbf{99}, 036602 (2007).

\bibitem{eto04}M. Eto, T. Ashiwa, and M. Murata. Current-induced
entanglement of nuclear spins in quantum dots. J. Phys. Soc. Jpn.
\textbf{73}, 307 (2004). 

\bibitem{christ07}H. Christ, J. I. Cirac, and G. Giedke. Quantum
description of nuclear spin cooling in a quantum dot. Phys. Rev. B
\textbf{75}, 155324 (2007).

\bibitem{dicke54}R. H. Dicke. Coherence in spontaneous radiation
processes. Phys. Rev. \textbf{93}, 99 (1954).

\bibitem{gross82}M. Gross and S. Haroche. Superradiance: An essay
on the theory of collective spontaneous emission. Phys. Rep. \textbf{93},
301 (1982).

\bibitem{kessler10}E. M. Kessler, S. Yelin, M. D. Lukin, J. I. Cirac,
and G. Giedke. Optical superradiance from nuclear spin environment
of single-photon emitters. Phys. Rev. Lett. \textbf{104}, 143601 (2010). 

\bibitem{schliemann03}J. Schliemann, A. Khaetskii, and Daniel Loss.
Electron spin dynamics in quantum dots and related nanostructures
due to hyperfine interaction with nuclei. J. Phys. Condens. Matter
\textbf{15}, R1809 (2003). 

\bibitem{taylor07}J. M. Taylor, J. R. Petta, A. C. Johnson, A. Yacoby,
C. M. Marcus, and M. D. Lukin. Relaxation, dephasing, and quantum
control of electron spins in double quantum dots. Phys. Rev. B \textbf{76},
035315 (2007). 

\bibitem{bracker05}A. Bracker, E. A. Stinaff, D. Gammon, M. E. Ware, J. G. Tischler, A. Shabaev, Al. L. Efros, D. Park,
D. Gershoni, V. L. Korenev, and I. A. Merkulov. 
Optical pumping of the electronic and nuclear spin of single charge-tunable quantum dots.
Phys. Rev. Lett. \textbf{94}, 047402 (2005). 

\bibitem{paget82}D. Paget. Optical detection of NMR in high-purity
GaAs: Direct study of the relaxation of nuclei close to shallow donors.
Phys. Rev. B \textbf{25}, 4444 (1982). 

\bibitem{ota07}T. Ota, G. Yusa, N. Kumada, S. Miyashita, T. Fujisawa,
and Y. Hirayama. Decoherence of nuclear spins due to dipole-dipole
interactions probed by resistively detected nuclear magnetic resonance.
Appl. Phys. Lett. \textbf{91}, 193101 (2007). 

\bibitem{bluhm10}H. Bluhm, S. Foletti, I. Neder, M. Rudner, D. Mahalu,
V. Umansky, and A. Yacoby. Dephasing time of GaAs electron-spin qubits
coupled to a nuclear bath exceeding 200 $\text{\ensuremath{\mu}}$s.
Nature Physics \textbf{7}, 109 (2010).

\bibitem{recher00}P. Recher, E. V. Sukhorukov, and Daniel Loss. Quantum
dot as spin filter and spin memory. Phys. Rev. Lett. \textbf{85},
1962 (2000). 

\bibitem{takahashi11}R. Takahashi, K. Kono, S. Tarucha, and K. Ono. 
Voltage-selective bidirectional polarization and coherent rotation
of nuclear spins in quantum dots. Phys. Rev.
Lett. \textbf{107}, 026602 (2011).

\bibitem{hanson04}R. Hanson, L.M.K. Vandersypen, L.H. Willems van
Beveren, J.M. Elzerman, I.T. Vink, and L.P. Kouwenhoven. Semiconductor
few-electron quantum dot operated as bipolar spin filter. Phys. Rev.
B \textbf{70}, 241304(R) (2004).

\bibitem{hanson03}R. Hanson, B. Witkamp, L.M.K. Vandersypen, L.H. Willems van
Beveren, J.M. Elzerman, and L.P. Kouwenhoven. Zeeman energy
and spin relaxation in one-electron quantum dot. Phys. Rev.
Lett. \textbf{91}, 196802 (2003).

\bibitem{agrawal71}G. S. Agrawal. Master-Equation approach to spontaneous
emission. III. Many-Body aspects of emission from two-level atoms
and the effect of inhomogeneous broadening. Phys. Rev. A \textbf{4},
1791 (1971).

\bibitem{leonardi82}C. Leonardi and A. Vaglica. Superradiance and
inhomogeneous broadening. II: Spontaneous emission by many slightly
detuned sources. Nuovo Cimento Soc. Ital. Fis. B \textbf{67}, 256
(1982).

\bibitem{temnov05}V. V. Temnov and U. Woggon. Superradiance and subradiance
in an inhomogeneously broadened ensemble of two-level systems coupled
to a low-Q cavity. Phys. Rev. Lett. \textbf{95}, 243602 (2005). 


\bibitem{gurvitz96}S. A. Gurvitz and Ya. S. Prager. Microscopic derivation
of rate equations for quantum transport. Phys. Rev. B \textbf{53},
15932 (1996). 

\bibitem{gurvitz98}S. A. Gurvitz. Rate equations for quantum transport
in multidot systems. Phys. Rev. B \textbf{57}, 6602 (1998). 

\bibitem{welack08}S. Welack, M. Esposito, U. Harbola, and S. Mukamel.
Interference effects in the counting statistics of electron transfers
through a double quantum dot. Phys. Rev. B \textbf{77}, 195315 (2008).

\bibitem{engel02}H.-A. Engel and D. Loss. Single-spin dynamics and
decoherence in a quantum dot via charge transport. Phys. Rev. B \textbf{65},
195321 (2002). 

\bibitem{timm08}C. Timm. Tunneling through molecules and quantum
dots: Master-equation approaches. Phys. Rev. B \textbf{77}, 195416
(2008). 

\bibitem{harbola06}U. Harbola, M. Esposito, and S. Mukamel. Quantum
master equation for electron transport through quantum dots and single
molecules. Phys. Rev. B \textbf{74}, 235309 (2006). 

\bibitem{zhao11}N. Zhao, J.-L. Zhu, R.-B. Liu, and C. P. Sun. Quantum
noise theory for quantum transport through nanostructures. New Journal
of Physics \textbf{13}, 013005 (2011).

\bibitem{yamamoto99}Y. Yamamoto and A. Imamoglu, 
\textit{Mesoscopic Quantum Optics} (Wiley, New York, 1999).

\bibitem{cohen-tannoudji92}C. Cohen-Tannoudji, J. Dupont-Roc, and G. Grynberg, 
\textit{Atom-Photon Interactions: Basic Processes and Applications}
(Wiley, New York, 1992).

\bibitem{timm12}C. Timm (private communication).

\bibitem{bruus06}H. Bruus and K. Flensberg, 
\textit{Many-Body Quantum Theory in Condensed Matter Physics} 
(Oxford University Press, New York, 2006).

\bibitem{cirac92}J. I. Cirac, R. Blatt, P. Zoller, and W. D. Phillips.
Laser cooling of trapped ions in a standing wave. Phys. Rev. A \textbf{46},
2668 (1992). 

\bibitem{bagrets03}D. A. Bagrets, and Yu. V. Nazarov. Full counting
statistics of charge transfer in Coulomb blockade systems. Phys. Rev.
B \textbf{67}, 085316 (2003). 

\bibitem{esposito09}M. Esposito, U. Harbola, and S. Mukamel. Nonequilibrium
fluctuations, fluctuation theorems, and counting statistics in quantum
systems. Rev. Mod. Phys. \textbf{81}, 1665 (2009). 

\bibitem{emary12}C. Emary, C. P\"oltl, A. Carmele, J. Kabuss, A. Knorr,
and T. Brandes. Bunching and antibunching in electronic transport.
Phys. Rev. B \textbf{85}, 165417 (2012). 

\bibitem{carmichael99}H. J. Carmichael, 
\textit{Statistical Methods in Quantum Optics 1}
(Springer, Berlin, 1999).

\bibitem{finitepol}For finite polarization the initial covariance matrix 
has been determined heuristically from 
the dark state condition $\left<A^{-}A^{+}\right>=0$
in the homogeneous limit.

\bibitem{binomial}This limit is realized if strong nuclear dephasing processes prevent the coherence build-up of the SR evolution.

\bibitem{carmichael80}H. J. Carmichael.
Analytical and numerical results for the steady state in cooperative resonance fluorescence. 
J. Phys. B \textbf{13}, 3551 (1980).

\bibitem{morrison08}S. Morrison and A. S. Parkins.
Collective spin systems in dispersive optical cavity QED: Quantum phase transitions and entanglement. 
Phys. Rev. A \textbf{77}, 043810 (2008).

\bibitem{kessler12} E. M. Kessler, G. Giedke, A. Imamoglu, S. F. Yelin, M. D. Lukin, and J. I. Cirac.
Dissipative phase transition in central spin systems.
arxiv: 1205.3341.

\bibitem{chung09}C.-H.Chung, K. Le Hur, M. Vojta, and P. W\"olfle.
Nonequilibrium transport at a dissipative quantum phase transition.
Phys. Rev. Lett. \textbf{102}, 216803 (2009).

\bibitem{borda06} L. Borda, G. Zarand, and D. Goldhaber-Gordon.
Dissipative quantum phase transition in a quantum dot.
arxiv: cond-mat/0602019.  

\bibitem{legett87}A. J. Leggett, S. Chakravarty, A. T. Dorsey, M. P. A. Fisher, A. Garg, and W. Zwerger.
Dynamics of the dissipative two-state system. 
Rev. Mod. Phys. \textbf{59}, 1 (1987).

\bibitem{rudner10}M. S. Rudner and L. S. Levitov.
Phase transitions in dissipative quantum transport and mesoscopic nuclear spin pumping.
Phys. Rev. B \textbf{82}, 155418 (2010). 
\end{thebibliography}
\end{document}